\documentclass[12pt]{article}
\pdfoutput=1
\usepackage[a4paper]{geometry}
\usepackage{jheppub, amsmath,amssymb,amsfonts,amsxtra, mathrsfs, makeidx,graphics,graphicx,amsthm,epsfig, bm,longtable,float, color,tikz,mathtools,xfrac,footnote,rotating, lscape, makecell, environ,mathtools, empheq}

\usepackage{subfig}
\usepackage{multirow}
\usepackage{adjustbox}

\usepackage{amsthm}
\usepackage{hyperref}

\usepackage{amsmath}
\usepackage{amsfonts}
\usepackage{commath}
\usepackage[english]{babel}
\usepackage{setspace}

\usepackage{epsfig,amssymb} 
\usepackage{amsmath}

\usepackage{amscd,color}
\usepackage{amsmath,graphicx}
\usepackage{verbatim,booktabs}

\usepackage{latexsym}
\usepackage{amsmath,amsfonts,amssymb,amsthm}
\usepackage{amsmath,amsthm}

\title{
\begin{center}
3d $\mathcal{N}=2$ dualities for $SU(N_c)\times U(1)$ Chern-Simons gauge theories
\end{center}
}

\author[a]{Antonio Amariti}
\author[a,b]{and Simone Rota}
\affiliation[a]{INFN, Sezione di Milano, Via Celoria 16, I-20133 Milano, Italy}
\affiliation[b]{Dipartimento di Fisica, Universit\`a degli Studi di Milano, Via Celoria 16, I-20133 Milano, Italy}
\emailAdd{antonio.amariti@mi.infn.it,  simone.rota@mi.infn.it}

\abstract{We systematically study 3d $\mathcal{N}=2$ dualities for $U(N_c)$ gauge theories
with different CS levels for the abelian and the non-abelian factors. We derive such dualities by
a gauging/ungauging procedure on other known dualities and by adding an opportune CS contact term for the baryonic symmetry. This allows us to follow the various steps on the three sphere partition function, obtaining exact identities among the new dual phases proposed here. We apply the procedure to various cases, namely SQCD without and with adjoint, chiral dualities and some sporadic cases arising from the real mass flow of models with an $E_7 \times U(1)$ enhancement of the global symmetry.}

\begin{document}

\maketitle

\section{Introduction}
\label{sec:intro}

A powerful and fascinating aspect of supersymmetric field theories is that they 
are a natural playground for the analysis of non-perturbative phenomena.
A central role in this context is played by the idea of duality. Supersymmetric dualities indeed map 
two (or more) field theories, described in terms of different degrees of freedom but sharing the same 
physical observables. Often such dualities are of strong/weak type and they allow for a perturbative access to the physics even if the starting description is in a non-perturbative regime.

A very active field of research in the last decade has been the case of 
dualities in 2+1 dimensions with four supercharges. The prototypical examples have been worked out in the late 90's in \cite{Aharony:1997bx,Aharony:1997gp,deBoer:1997ka} and more recently in 
\cite{Aharony:2008gk,Giveon:2008zn}, including the presence of CS terms as well.
Many tests and generalizations of these dualities have then been possible thanks to the results from localization on the three sphere \cite{Kapustin:2009kz,Jafferis:2010un} and from the computation of the superconformal index \cite{Kim:2009wb} and of the twisted index \cite{Benini:2015noa}.
Such supersymmetric dualities have then inspired the search of analogous non-supersymmetric ones, and many examples have been recently worked out after the seminal papers  \cite{Aharony:2015mjs,Seiberg:2016gmd,Karch:2016sxi}.
Many of the results  obtained in the non-supersymmetric case indeed resemble the ones derived in the supersymmetric one. For some models a derivation of the non-supersymmetric cases  
starting from the supersymmetric one has been furnished in the literature (see for example \cite{Gur-Ari:2015pca,Kachru:2016rui,Kachru:2016aon}).

This similarity suggests   to re-consider the web of three dimensional supersymmetric dualities and to look for the cases that share the same behavior of 
non-supersymmetric cases.
Indeed, even in the simplest cases with a single $U(N_c)$ gauge group, we have  only partial understanding of supersymmetric dualities with unitary gauge groups and a different CS level for the $SU(N_c)$ and for the $U(1)$ factor.
Only very recently in \cite{Nii:2020ikd} a duality of this type has been proposed for the case of three dimensional SQCD with non-vanishing CS level for the non-abelian factor. On the other hand in the non-supersymmetric cases these types of dualities have been studied more deeply, se for example  \cite{Hsin:2016blu,Radicevic:2016wqn}.

Motivated by this discussion in this paper we study 3d $\mathcal{N}=2$ dualities that involve 
$U(N_c)_{k_1, k_1+N_c k_2} \equiv (
SU(N_c)_{k_1} \times U(1)_{N_c(k_1+N_c k_2)})/\mathbb{Z}_{N_c}$ gauge groups. We show how the dualities can be constructed starting from 
well established dualities (e.g. Aharony duality \cite{Aharony:1997gp})  by a dynamical mechanism that mixes real mass flow, gauging and ungauging of abelian symmetries and  coupling to topological sectors.
By applying this procedure we first recover the generalized Giveon-Kutasov dualities obtained in \cite{Nii:2020ikd} and then we generalize the construction to many other  3d $\mathcal{N}=2$ dualities.
We then show that our procedure can   be applied to the integral identities relating the supersymmetric three sphere partition functions of the dual phases.
This is a necessary and non-trivial check of the new dualities that we are proposing here.

The paper is organized as follows. 
In section \ref{FTsec} we discuss the general prescription that we adopt in the rest of the paper for constructing the dualities for $U(N_c)_{k_1,k_1+N_c k_2}$ starting from the ones with 
$U(N_c)_{k_1} \equiv U(N_c)_{k_1,k_1}$ gauge groups.
In section \ref{sec:new} we apply the prescription to the case of Aharony duality finding a new duality for $U(N_c)_{0,N_c k_2}$ and we corroborate our proposal by matching the three sphere partition function.
Then in section \ref{sec:genGK} we derive the generalized Giveon-Kutasov duality for 
$U(N_c)_{k_1,k_1+N_c k_2}$  proposed in \cite{Nii:2020ikd}.
In section \ref{sec:adj} we consider the case of $U(N_c)$ adjoint SQCD with a power law superpotential
discussing both the case with vanishing and the case with non-vanishing CS level for the non-abelian factor.
In section \ref{sec:chiral} we discuss the case of SQCD without and with an adjoint field and chiral matter content, i.e. a different amount of fundamentals and anti-fundamentals.
In section \ref{sec:ex} we consider a triality proposed in  \cite{Amariti:2018wht,Benvenuti:2018bav} involving $U(N_c)_0$ with two flavors and adjoint without a power law superpotential.
In section \ref{sec:conc} we conclude our analysis by commenting on possible future directions.

%
%
%
%
%
%
\section{Field theory}
\label{FTsec}
%
%
%
%
%
%

In this section we discuss the field theory construction that we use in order to 
modify models with $U(N_c)_{k_1}$ gauge group  and obtain models with $U(N_c)_{k_1,k_1+N_c k_2}$ gauge group.
The construction is inspired by the so called gauging/ungauging procedure \cite{Aharony:2013dha},  that connects $SU(N_c)$ and $U(N_c)$ SQCD in three dimensions.
In the gauging procedure one considers an $SU(N_c)$ theory with a baryonic symmetry $U(1)_{x}$, assigning charge $\frac{1}{N_c}$ to the fundamentals and $-\frac{1}{N_c}$ to the anti-fundamentals.
Then by making dynamical the corresponding background gauge field associated to $U(1)_{x}$ the gauge group becomes  $U(N_c) = SU(N_c) \times U(1)/\mathbb{Z}_{N_c}$.
In this theory an extra symmetry emerges, due to the Bianchi identity, corresponding to the topological $U(1)_J$ symmetry. The real mass associated to such a new symmetry is the FI term for the dynamical $U(1)$ gauge field.
The ungauging procedure corresponds to the reverse operation, starting from $U(N_c)$ and  leading to $SU(N_c)$.
This operation consists of considering a background gauge field associated to the $U(1)_J$ symmetry and then making this gauge field dynamical.
In this case we gain a mixed CS term, corresponding to the original FI,  between $U(1)_J$ and
$U(1) \subset U(N_c)$. There is also a  further topological $U(1)$ (that we can denote as $U(1)_{J'}$), associated to the gauged $U(1)_J$.
The mixed CS makes the two abelian gauge fields massive and one can integrate them out
(if there are no matter fields charged under these symmetries).
The gauge group becomes $SU(N_c)$ and the new topological symmetry 
$U(1)_{J'}$ corresponds to the baryonic symmetry $U(1)_{x}$, normalized such that the fundamentals have baryonic charge $\frac{1}{N_c}$.
Summarizing, if we apply the ungauging procedure to an $U(N_c)$ gauge group we obtain 
 $SU(N_c)$, while we can come  come back to the original $U(N_c)$ gauge group by the gauging procedure.

In the following we modify this process with a further step: it corresponds to the addition of an extra global CS 
term for the baryonic symmetry in the $SU(N_c)$ case. This becomes a CS term when we gauge the baryonic symmetry in order to recover the $U(N_c)$ gauge theory.
The final theory is $U(N_c)_{0,N_c k_2}$.
If we further consider SQCD with $N_f+k_1$ pairs of fundamentals and anti-fundamentals and give a large (positive) real mass to $k_1$ pairs we then end up with  $U(N_c)_{k_1,k_1+N_c k_2}$ SQCD with $N_f$ pairs of fundamentals and anti-fundamentals. 

This  procedure can be used in order to construct new dualities starting from known ones. It is necessary for its applicability that we are in absence of (monopole) superpotentials that break the topological symmetry of the original $U(N_c)$ gauge theory.

Furthermore in the dual phase there are often  fields charged under the topological symmetry.
They usually corresponds to singlets of the dual phases associated to monopoles of the original theory and they  are necessary to match the moduli space among the dual phases.
The presence of such fields  makes the analysis more complicated as we will see in the examples below. 
Indeed in such cases it is not possible in general to integrate out the massive vector multiplets for the abelian  gauge symmetry, even if in some cases one can take advantage of  supersymmetric dualities, to be used \emph{locally} (i.e. in a sub-sector of the model), 
in order to simplify the dual phase.
Furthermore depending on the matter content charged under $U(1)_J$ there can be   extra global CS  terms that can   obstruct the gauging of $U(1)_J$ if not properly quantized.
We will discuss these various issues and possibilities in the examples below.

A natural playground to apply this procedure is the three sphere partition function. Indeed gauging a symmetry corresponds to an integral over a real mass parameter and CS terms are gaussian insertions in the integrals.
In the following we will derive the identities for the new dualities involving 
$U(N_c)_{k_1,k_1+N_c k_2}$ (starting from the case with $k_1=0$)  gauge groups by applying our procedure on the integral identities 
involving the original  $U(N_c)_{0}$ cases and then we will study the case with non-zero CS level for the non-abelian factor by real mass flow on the fundamental fields.
Anyway, as we will comment below, it is  possible to start from integral identities with a non-zero CS level for the non-abelian factor as well. In these cases we must pay attention to correctly identify the topological sectors involving the CS levels and integrate over them in the partition function.

%
%
%
%
%
%
\section{ A new duality for $U(N_c)_{0,\kappa}$ SQCD}
\label{sec:new}
%
%
%
%
%
%
We start our survey with the case of SQCD with fundamental matter.
The strategy adopted here is the following: we start by considering Aharony duality \cite{Aharony:1997gp}. 
The electric theory is
$U(N_c)_{0}$ SQCD with $N_f$ 
pairs of fundamentals $Q$ and  anti-fundamentals $\tilde Q$
and vanishing superpotential.
The magnetic theory is $U(N_f-N_c)_{0}$ SQCD with $N_f$ 
pairs of dual fundamentals $q$ and  anti-fundamentals $\tilde q$,
$N_f^2$ singlets $M$ corresponding to the meson $Q \tilde Q$ and two extra singlets $T$ and $\tilde T$ identified with the 
monopole and anti-monopole operators of the electric phase. The dual theory has also 
a non-vanishing superpotential 
\begin{equation}
\label{WdAh}
W = M q \tilde q +  T t + \tilde T \tilde t
\end{equation}
where $t$ and $\tilde t$ are the monopole and anti-monopole operators of the dual theory.
The fields and the monopole operators have the following charges under the global 
$SU(N_f)^2 \times U(1)_A \times U(1)_J \times U(1)_R$
symmetries
\begin{equation}
\begin{array} {c|ccccc}
             & SU(N_f)_L & SU(N_f)_R &U(1)_A & U(1)_J & U(1)_R \\
             \hline
Q          &     N_f    &      1   & 1 & 0  & \Delta \\
\tilde Q  &    1        &   \overline N_f   & 1 & 0 & \Delta \\
q            &    1        &   N_f   & - 1 & 0 &1-\Delta \\
\tilde q   &     \overline N_f         &   1   & - 1 & 0 & 1-\Delta\\
M           &     N_f    &       \overline N_f     & 2 & 0 &2\Delta \\
T            &     1    &       1     &  N_f & 1 & N_f (1-\Delta) -N_c+1 \\
\tilde T   &     1    &       1     &  N_f & - 1 &  N_f (1-\Delta) -N_c+1\\
t            &     1    &       1     &  -N_f & -1 & N_f (\Delta-1) +N_c+1 \\
\tilde t     &     1    &       1     &  -N_f & 1 &  N_f (\Delta-1)+N_c+1 
\end{array}
\end{equation}

The next step consists of gauging the topological $U(1)_J$ symmetry (this is the ungauging procedure explained above).
The gauge group becomes $U(N_c) \times U(1)_J$, with a mixed CS at level $1$ between the two gauge abelian factors.
There is also a topological symmetry introduced by this procedure for the $U(1)_J$ symmetry, that corresponds to the
baryonic symmetry for the $SU(N_c)$ gauge group, and that we denote as $U(1)_{x}$. The fundamentals have charge $\frac{1}{N_c}$ under this symmetry.
We then couple the theory to a background CS at level $\kappa$ for the $U(1)_{x}$ symmetry and then we gauge the 
$U(1)_{x}$ symmetry as well.
Then we integrate out the massive vector fields associated to $U(1) \subset U(N_c)$ and to $U(1)_J$. On the other hand the gauged baryonic $U(1)_{x}$ symmetry cannot be integrated out because the fundamentals and anti-fundamentals are charged under it. This $U(1)_{x}$ indeed corresponds to the $U(1)$ factor for the $U(N_c)$ gauge group . The difference between the original electric theory and the new theory that we have just constructed is that the $U(1) \subset U(N_c)$ factor has CS level 
$N_c^2 \kappa$. The final gauge group is then 
$$
U(N_c)_{0,N_c \kappa}
=
\frac{SU(N_c)_{0} \times U(1)_{N_c^2 \kappa}}{\mathbb{Z}_{N_c}}
$$

This  construction  produces also a  dual phase for the electric 
$U(N_c)_{0,N_c \kappa}$ theory constructed so far.
We can indeed follow the same steps discussed above for the magnetic $U(N_f-N_c)_0$ 
gauge theory. In this case there are singlets charged under $U(1)_J$ and the massive gauge fields,  due to the mixed CS between $U(1) \subset U(N_f-N_c)$ and $U(1)_J$, cannot be integrated out.
The dual theory has then $U(N_f-N_c) \times U(1)_J \times U(1)_{x}$ gauge group with non-trivial (mixed) 
CS terms for the abelian factors.
We can further simplify this model by using a local mirror symmetry. Indeed  there are only two fields charged under $U(1)_J$, with charges
$\pm 1$, i.e. this theory corresponds to SQED and it is mirror dual to the XYZ model.  The topological symmetry for $U(1)_J$ is 
$U(1)_{x}$ (in this case the fundamentals and the anti-fundamentals have charge $\frac{1}{\tilde N_c}$ under it) and it implies that in the dual XYZ model the baryon and the anti-baryon are associated to the fields
$Y$ and $Z$. The further singlet $X$ has   the same quantum numbers of the monopole operators of the 
electric $SU(N_c)$ theory. 
All in all we have found a duality between 
\begin{itemize}
\item An electric $U(N_c)_{0,N_c \kappa}$ gauge theory with $N_f$ pairs of fundamentals $Q$ and anti-fundamentals 
$\tilde Q$ with $W=0$
\item A magnetic
 $U(N_f-N_c)_0 \times U(1)_{\kappa}$ gauge theory\footnote{Where the normalization for the CS term of the abelian gauge group misses the extra $N_c^2$ factor because of the different normalization of the fundamentals.}
 with  $N_f$ pairs of fundamentals $q$ and anti-fundamentals 
$\tilde q$, the meson $M$, the baryons $B$ and $\tilde B$ and another singlet $X$, corresponding to the monopole of the electric $SU(N_c)_0$ gauge group. The superpotential of the dual theory is
$$
W = M q \tilde q + X B \tilde B
$$
\end{itemize}

\subsection{Integral identities from the three sphere partition function}

We can reproduce the discussion above on the (squashed) three sphere partition function. Our strategy consists of
deforming the integral identity  for the Aharony duality and then 
obtaining the new identity for the new duality between the $U(N_c)_{0, N_c \kappa}$ and the 
$U(N_f-N_c)_0 \times U(1)_{\kappa}$ models discussed above.

The Aharony duality is reproduced on the three sphere by the identity 
\begin{align}
		Z_\text{ele}^{U(N_c)} 	(\mu;\nu;\lambda) =&
		\prod_{a,b=1}^{N_f}
			\Gamma_h(\mu_a + \nu_b)
			Z_\text{mag}^{U(N_f - N_c)}	(\omega - \nu;\omega - \mu;-\lambda)
		\nonumber\\
		&
		\Gamma_h	\left(
						\pm \frac{\lambda}{2} - m_A N_f 
						+ \omega(N_f + N_c + 1)
					\right)\ ,
	\end{align}
where
	\begin{equation}
		Z^{U(N_c)} (\mu;\nu;\lambda) =
		\int
		\prod_{i=1}^{N_c}	d\sigma_i 	e^{i\pi \lambda \sum_i \sigma_i}
		\prod_{a=1}^{N_f}	\Gamma_h(\mu_a + \sigma_i)	\Gamma_h(\nu_a - \sigma_i)
		\prod_{1 \leq i < j \leq N_c}	\Gamma_h^{-1} 	(\pm (\sigma_i - \sigma_j))\ .
	\end{equation}
The functions $\Gamma_h$ are hyperbolic Gamma functions (see for example \cite{VanDeBult} for definitions) and they correspond to the one loop
determinants obtained by localizing the action on the curved manifold.

In order to get an $SU(N_c)_0$ theory on the electric side we add a term $\frac{1}{2} e^{-i\pi\lambda N_c m_B}$ and gauge the topological symmetry by integrating over $\lambda$:	
	\begin{align}	\label{eq:SU(N)0_FI}
		Z_\text{ele} =& 
		\frac{1}{2}
		\int d\lambda
		\int \prod_{i=1}^{N_c} 	d \sigma_i 	\,
			\frac{e^{i\pi \lambda \sum_i (\sigma_i - m_B)}}{\prod_{i < j}^{N_c} \Gamma_h (\pm (\sigma_i - \sigma_j))} 
		\prod_{a=1}^{N_f}
		\Gamma_h(\mu_a + \sigma_i)
		\Gamma_h(\nu_a - \sigma_i) \ .
	\end{align}	
On the magnetic side we get, defining $\lambda = 2\xi$ (as in \cite[Eq. (3.7)]{Amariti:2020xqm}) and $\tilde{N}_c = N_f-N_c$:
	\begin{align}	\label{eq:SU(N)0_mag}
		Z_\text{mag} =&
		\prod_{a,b=1}^{N_f}	\Gamma_h(\mu_a + \nu_b)
		\int d\xi \,\Gamma_h	\left(	
								\pm \xi - m_A N_f + \omega(\tilde{N}_c + 1)
							\right)
		\nonumber\\
		&\int  \prod_{i=1}^{\tilde{N}_c} 	d \tilde{\sigma}_i 
			\prod_{i \leq i < j \leq \tilde{N}_c} \Gamma_h (\pm (\tilde{\sigma}_i - \tilde{\sigma}_j))^{-1}
			e^{-2 \pi i \xi 
				\left(
					\sum_i (\tilde{\sigma}_i + (N_c/\tilde{N}_c)	m_B)
				\right)
			}
		\nonumber\\
		& \prod_{i=1}^{\tilde{N}_c}	\left(
			\prod_{a=1}^{N_f}	\Gamma_h(\omega - \mu_a -\tilde{\sigma}_i)
								\Gamma_h(\omega - \nu_a +\tilde{\sigma}_i)
		\right)\ .
	\end{align}
Now, starting from \eqref{eq:SU(N)0_FI} we add a contact term
 $e^{-i\pi (\kappa x^2-2 \Lambda x)}$ 
 and we gauge the baryonic symmetry $x=m_B N_c$:
	\begin{equation}
		\int dx e^{-\pi i \kappa x^2 + 2\pi i x(\Lambda - \xi)}
		=
		e^{\frac{\pi i}{\kappa}	(\Lambda - \xi)^2}\ .
	\end{equation}	
The $d\xi$ ($= \frac{1}{2}d\lambda$) integration then gives:
	\begin{align}
		\int  d \xi
			e^{  \frac{\pi i}{\kappa}	(\Lambda^2 + \xi^2)
				+ 2\pi i \xi \left(  \sum_i 	\sigma_i -\frac{\Lambda}{\kappa}\right) 
			}
		=
		e^{
			-\pi i\kappa
			\left( \sum_i\sigma_i -\frac{\Lambda}{\kappa}	\right)^2
			+
			\frac{\pi i}{\kappa}	\Lambda^2
		}
		=
		e^{
			-i\pi\kappa
			\left( \sum_i\sigma_i \right)^2
			+
			2\pi i \Lambda
			 \sum_i\sigma_i
		}\ .
	\end{align}
Therefore the full electric partition function becomes:
	\begin{align} \label{U(N)0k_ele}
		Z_\text{ele} =&
		\int \prod_{i=1}^{N_c} 	d \sigma_i 	
			e^{
			-\pi i\kappa	\left( \sum_i\sigma_i \right)^2
			+
			2\pi i \Lambda	\sum_i\sigma_i
			}
		\prod_{i \leq i < j \leq N_c} \Gamma_h (\pm (\sigma_i - \sigma_j))^{-1}
		\nonumber\\
		&
		\prod_{a=1}^{N_f}
		\Gamma_h(\mu_a + \sigma_i)
		\Gamma_h(\nu_a - \sigma_i)
	\end{align}
We recognize the partition function of $U(N_c)_{0,N_c\kappa}$ (with an FI term and flavors) according to \cite{Amariti:2020xqm}.  
We can perform the same steps in the magnetic theory  by starting from \eqref{eq:SU(N)0_mag}, then by gauging the baryonic symmetry and eventually by performing the $\xi$ integral. We obtain:
	\begin{align}	\label{U(N)0k_mag}
		Z_\text{mag} =&
		\prod_{a,b=1}^{N_f}	\Gamma_h (\mu_a + \nu_b)	
		\Gamma_h\left(
			2\omega(N_f - N_c + 1) - \sum_a (\mu_a + \nu_a)
		\right)
		\int dx e^{-\pi i  \kappa x^2 + 2\pi i x \Lambda}
		\nonumber\\
		&\int
		\frac{\prod_{i=1}^{\tilde{N}_c}	d\tilde{\sigma}_i }
			{\prod_{i< j}^{\tilde{N}_c} \Gamma_h (\pm(\tilde{\sigma}_i - \tilde{\sigma}_j)) }
			\Gamma_h\left(
				\pm \left(\sum_{i=1}^{\tilde{N}_c} \tilde{\sigma}_i + x \right)
				+ \frac{1}{2} \sum_a (\mu_a + \nu_a) 
				- \omega \tilde{N}_c
			\right)
		\nonumber\\
		& \prod_{i=1}^{\tilde{N}_c}	\left(
		\prod_{a=1}^{N_f}
		\Gamma_h	\left( 
				\omega - \mu_a -\tilde{\sigma}_i					
			\right)
		\Gamma_h	\left( 
				\omega - \nu_a +\tilde{\sigma}_i					
			\right)
		\right)
	\end{align}
that corresponds  to the partition function of the $U(N_f-N_c)_0 \times U(1)_{\kappa}$ theory.

This is a new duality that was not discussed in \cite{Nii:2020ikd}, where only theories with non-zero CS level for the non-abelian group were considered. 
The difference between the dualities discussed in \cite{Nii:2020ikd}, for $U(N_c)_{k_1,k_1+N_c k_2}$ gauge group, and the case discussed here, that corresponds to setting $k_1=0$, is that here we must consider further fields on the dual side.
These fields correspond in the Aharony duality to the monopoles of the electric theory that act as singlets in the dual phase, setting the monopole of the magnetic phase to zero in the chiral ring.
In the case studied here we have seen that after the gauging/ungauging procedure such fields are no longer singlets 
in the dual phase, but are charged under the "baryonic" $U(1)_x$  sector. This is the same phenomenon-already observed in the literature \cite{Aharony:2013dha,Park:2013wta} in the case of Aharony duality for $SU(N_c)_{0}$ SQCD.

%
%
%
%
\section{ The generalized GK duality for  $U(N_c)_{k_1,k_1+N_c k_2}$}
\label{sec:genGK}
%
%
%
%
%
%
 In this section we generalize the analysis to $U(N_c)_{k_1,k_1+N_c k_2}$ SQCD.
 This duality has been studied in \cite{Nii:2020ikd} and it has been named there as generalized Giveon-Kutasov
 duality.
 
 The simplest way to obtain this duality consists of considering the duality for $U(N_c)_{0,N_c \kappa}$
 (with $\kappa \equiv k_2$) and consider $N_f+k_1$ pairs of fundamentals and antifundamentals.
 The dual model has   gauge group $U(N_f+k_1-N_c)_0 \times U(1)_{k_2}$.
 The further step to obtain the generalized Giveon-Kutasov duality corresponds to assign 
 a large real (positive or negative) mass to $k_1$ fundamentals and antifundamentals.
 The electric theory becomes $U(N_c)_{\pm k_1,\pm k_1+N_c k_2}$ with $N_f$ pairs of fundamentals and antifundamentals. The dual theory has gauge group  $U(N_f+k_1-N_c)_{\mp k_1,\mp ( N_f -N_c)} \times U(1)_{k_2+1}$: 
 with a mixed CS term at level $-1$ between $U(1) \subset U(N_f+k_1-N_c)$ and the other $U(1)$ factor.
There are $N_f$ pairs of dual fundamentals $q$ and anti-fundamentals $\tilde q$, $N_f^2$ singlets $M$ identified with the mesons of the electric theory, and the superpotential is just $W = M q \tilde q$.

On the partition function we shift the masses as
	\begin{align}
		\mu_a \rightarrow \Bigg \lbrace
			\begin{split}
				&(m_a - \frac{k_1}{N_f + k_1}s) + (m_A + \frac{k_1}{N_f + k_1}s)
				\qquad	
				a = 1,\dots ,N_f
				\\
				&(m_a + \frac{N_f}{N_f + k_1}s) + (m_A + \frac{k_1}{N_f + k_1}s)
				\qquad
				a=N_f+1,\dots,N_f+k_1
			\end{split}
		\\
		\nu_a \rightarrow \Bigg \lbrace
		\begin{split}
			&(n_a - \frac{k_1}{N_f + k_1}s) + (m_A + \frac{k_1}{N_f + k_1}s)
			\qquad	
			a = 1,\dots ,N_f
			\\
			&(n_a + \frac{N_f}{N_f + k_1}s) + (m_A + \frac{k_1}{N_f + k_1}s)
			\qquad
			a=N_f+1,\dots,N_f+k_1
		\end{split}\ .
	\end{align}
and we study the limit $|s| \rightarrow \infty$ in the integral identity between (\ref{U(N)0k_ele}) and (\ref{U(N)0k_mag}).

After removing the divergent contributions, that  we have shown to match in the electric and in the magnetic phases, we are left with the identity between (for $s>0$, the case $s<0$ can be studied analogously):
\begin{align}	\label{eq:U(N)k1k2_ele}
	Z_\text{ele} =&\frac{1}{N_c!}
	\int \prod_{i=1}^{N_c} 	d \sigma_i 	
	e^{
		-\pi i k_2	\left( \sum_i\sigma_i \right)^2
		-
		\pi i k_1 \sum_i \sigma_i^2
		+
		2\pi i	\sum_i\sigma_i \left(\Lambda + \frac{1}{2}\sum_{a=1}^{N_f} (n_a - m_a)	\right)
	}	
\nonumber\\
&
	\prod_{i \leq i < j \leq N_c} \Gamma_h (\pm (\sigma_i - \sigma_j))^{-1}
	\prod_{a=1}^{N_f}
	\Gamma_h(\mu_a + \sigma_i)
	\Gamma_h(\nu_a - \sigma_i)
\end{align}
and  
\begin{align}	\label{eq:U(N)k1k2_mag}
	Z_\text{mag} =&\frac{e^{-\frac{i\pi \phi}{2}}}{\tilde{N}_c!}
	\prod_{a,b=1}^{N_f}	\Gamma_h (\mu_a + \nu_b)	
	\int dx 
	\int \prod_{i=1}^{\tilde{N}_c}	d\tilde{\sigma}_i 
	e^{\pi i \left( -(k_2+1) x^2 + 2 x \left( \Lambda + \sum_i \tilde{\sigma}_i \right)\right)}
\nonumber\\	&
	e^{	\pi i  \left( k_1\sum_i \tilde{\sigma}_i^2- \left(\sum_i \tilde{\sigma}_i \right)^2\right)}
	e^{	\pi i  \sum_i \tilde{\sigma}_i	\sum_{a=1}^{N_f} (m_a - n_a)}
\nonumber\\&
	 \prod_{i=1}^{\tilde{N}_c}
	\prod_{a=1}^{N_f}
	\frac{	\Gamma_h	\left( \omega - \mu_a -\tilde{\sigma}_i		\right)
		\Gamma_h	\left(  \omega - \nu_a +\tilde{\sigma}_i	\right)}
		{\prod_{i< j}^{\tilde{N}_c} \Gamma_h (\pm(\tilde{\sigma}_i - \tilde{\sigma}_j))}
\end{align}
With
\begin{align}	\label{eq:U(N)k1k2_ct}
	\phi =  &-k_1 \left(\sum _{a=1}^{N_f} m_a^2+\sum _{a=1}^{N_f} n_a^2\right)-2 \omega  N_c \left(2 N_f \left(\omega -m_A\right)+k_1 \omega \right)
\nonumber\\&
	+2 k_1 N_f \left(\omega ^2-m_A^2\right)+2 N_f^2 \left(\omega -m_A\right){}^2+2 \omega ^2 N_c^2+\left(k_1^2-1\right) \omega ^2
\end{align}%
and $\tilde N_c = N_f+k_1-N_c$.
Observe that the CS level for the abelian factor inside $U(\widetilde N_c)$ in the dual partition function can be read by summing up the two contributions coming from $\sum_i \sigma_i^2$ and $(\sum_i \sigma_i)^2$.
In the first case the level is $-k_1$ while in the second case it is $N_f-N_c+k_1$, and the final contribution corresponds to 
$U(1)_{N_f-N_c}$ as expected (for $s<0$ the corresponding contribution is $U(1)_{N_c-N_f}$).

The equivalence between (\ref{eq:U(N)k1k2_ele}) and (\ref{eq:U(N)k1k2_mag}) corresponds to the identity of the partition function for the generalized 
Giveon-Kutasov duality introduced in \cite{Nii:2020ikd} as discussed above.

The dualities discussed in sections \ref{sec:new} and \ref{sec:genGK} are between a theory with gauge group $U(N_c)$, which we called the electric theory, and a theory with gauge group $U(\tilde{N}_c)\times U(1)$, called the magnetic theory. The magnetic gauge group is originally $U(\tilde{N}_c)\times U(1) \times U(1)$, where the two abelian factors come from the gauging of the electric topological symmetry $U(1)_J$ and of the baryonic symmetry $U(1)_x$. One of the two abelian factors can be eliminated by a local mirror duality, which corresponds to performing the $\xi$ integral in the magnetic partition function. The remaining abelian sector in the magnetic theory cannot be integrated out in the same way because performing the gaussian $x$ integral in  \eqref{eq:U(N)k1k2_mag} results in a fractional CS coefficient for the magnetic $U(N_f - N_c)$ gauge group, which breaks  the gauge symmetry.

On the other hand there are instances, among the duality considered in this paper, where such a mirror symmetry cannot be performed, or does not simplify the magnetic gauge group. 
In this paper we will allow for magnetic theories with multiple additional abelian gauge sectors, making sure to carefully account for the possible Chern Simons terms, mixed Chern Simons  terms, FI terms and charged matter fields for the additional gauge sectors. 
Moreover we notice that the duality for $U(N_c)_{k_1,k_1+N_c k_2}$ SQCD just derived via a real mass flow can be   obtained directly from Giveon-Kutasov duality as well. This is achieved by applying the same gauging/ungauging procedure described for the case of Aharony duality. The resulting duality coincides with the one explicitly obtained via real mass flow, therefore the infinite mass limit in the flow and the gauging/ungauging procedure commute for these theories.
Another consistency check of this duality is  that for $k_2=0$ the magnetic $U(1)$ gauge sector can be integrated away and the resulting duality is Giveon-Kutasov duality.
%
%
%
%
%
\section{Adjoint SQCD}
\label{sec:adj}
%
%
%
%
%
%
In this section we extend our analysis to $U(N_c)$ CS SQCD with adjoint matter.
The original duality in this case has been found in \cite{Kim:2013cma} and it relates
\begin{itemize}
\item
$U(N_c)_0$ SQCD with $N_f$ fundamentals and anti-fundamentals and an adjoint 
 $X$ with superpotential $W = Tr X^{n+1}$
 \item  
$U(n N_f - N_c)_0$ SQCD with $N_f$ dual fundamentals $q$ and anti-fundamentals $q$, an adjoint 
 $Y$,  $n N_f^2$ singlets $M_j = Q X^j \tilde Q$, for $j=0,\dots,n-1$ and $2n$ singlets $T_j$ and $\tilde T_j$.
 The dual  superpotential is
 \begin{equation}
W = Tr Y^{n+1} + \sum_{j=0}^{n-1} (M_j q Y^j \tilde{q} + t_j  T_{n-j}+ \tilde{t}_j \tilde{T}_{n-j})
 \end{equation}
 \end{itemize}
 Again we follow the procedure explained in section \ref{FTsec}, by first gauging the topological symmetry $U(1)_J$, then coupling the new topological symmetry $U(1)_{J'}$, arising from this gauging, to a topological sector and then by gauging $U(1)_{J'}$
 as well.
 The final duality that we obtain relates
 \begin{itemize}
\item
$U(N_c)_{0,N_c \kappa} $ SQCD with $N_f$ fundamentals and anti-fundamentals and an adjoint 
 $X$ with superpotential $W = Tr X^{n+1}$
 \item  A dual gauge theory with gauge group
 \begin{equation}
	\overbracket{U(\tilde{N}_c \equiv n N_f-N_c)_{0}	 \times U(1)_0}^{1}
	\overbracket{	\times 	U(1)_{\kappa}}^{1}
\end{equation}
 with $N_f$ dual fundamentals $q$ and anti-fundamentals $\tilde q$ and an adjoint 
 $Y$ of  $U(n N_f-N_c)$ gauge group,  $n N_f^2$ singlets $M_j = Q X^j \tilde Q$, for $j=0,\dots,n-1$ and $2n$ fields with charge $\pm 1$ under $U(1)_{0}$, that we denote as $V_j$ and $\tilde V_j$.
 The dual  superpotential is
 \begin{equation}
W = Tr Y^{n+1} + \sum_{j=0}^{n-1} (M_j q Y^j \tilde{q} + t_j  V_{n-j}+ \tilde{t}_j \tilde{V}_{n-j})
 \end{equation}
 where $t_j$ and $\tilde t_j$ are the dressed monopole operators of the non-abelian gauge group. Observe that in this case
 the effective FI for $U(n N_f - N_c)_0$ corresponds to a mixed CS term, i.e. these operators are charged under the abelian gauge group $U(1)_0$.
 \end{itemize}
The corresponding integral identity relating the three sphere partition functions of the duality of 
\cite{Kim:2013cma}  is $Z_{ele} = Z_{mag}$, where
$Z_{ele} = Z_{U(N_c)_0}(\mu,\nu;\tau;\xi)$ 
with
\begin{align}
\label{zuad}
	Z_{U(N_c)_k}  (\mu,\nu;\tau;\xi)=&
	\frac{\Gamma_h(\tau)^{N_c}}{N_c!}
	\int \prod_{i=1}^{N_c}	d\sigma_i e^{-\pi i (k \sigma_i^2-2  \xi  \sigma_i)}
		\prod_{i<j}	\frac{\Gamma_h(\tau \pm (\sigma_i-\sigma_j))}
						{\Gamma_h(\pm (\sigma_i-\sigma_j))}
\nonumber\\&
	\prod_{i=1}^{N_c} 
	\left(
		\prod_{a=1}^{N_f}
		\Gamma_h(\mu_a + \sigma_i)
		\Gamma_h(\nu_a - \sigma_i)
	\right)
\end{align}
and
\begin{align}
	Z_{\text{mag}} =&
	\prod_{j=0}^{n-1}
	\Gamma_h\left(
		\pm \xi
		-\frac{1}{2} \sum (\mu_a - \nu_a)
		+\tau (j-N_c+1)
		+\omega N_f
	\right)
\nonumber\\&
	\prod_{a,b=1}^{N_f} \Gamma_h(\mu_a + \nu_b + j\omega\tau)
	\frac{\Gamma_h(\tau)^{\tilde{N}_c}}{\tilde{N}_c!}
	 Z_{U(\widetilde N_c)_0}(\tau-\mu,\tau-\nu;\tau;-\xi)
\end{align}
Observe that in both $Z_{ele}$ and $Z_{mag}$ we fixed $\tau = \frac{2 \omega}{n+1}$ because of the adjoint superpotential.

We then apply the prescription that we have discussed above on the field theory dualiity directly on the integral identity. It corresponds to 
\begin{itemize}
	\item Multiply both partition functions by a factor $e^{-2\pi i \xi N_c m_B}$ and integrate in $d\xi$
	\item Multiply both partition functions by a factor $e^{-\pi i \kappa x^2 +  2 \pi i x \Lambda}$ and integrate in $dx$, with $x=m_B N_c$
\end{itemize}
On the electric side we obtain the partition function for  the $U(N)_{0, N_c \kappa}$ gauge theory with $N_f$ flavors in the fundamental and anti-fundamental and the adjoint:
\begin{align}
\label{eee}
	Z_{\text{ele}}  =&
	\frac{\Gamma_h(\tau)^{N_c}}{N_c!}
	\int \prod_{i=1}^{N_c}	d\sigma_i
	e^{
		-\pi i \kappa
		\left( \sum_i\sigma_i \right)^2
		+
		2\pi i \Lambda
		\sum_i\sigma_i
	}
\nonumber\\&
	\prod_{i<j}	\frac{\Gamma_h(\tau \pm (\sigma_i-\sigma_j))}
	{\Gamma_h(\pm (\sigma_i-\sigma_j))}
	\prod_{i=1}^{N_c} 
	\left(
	\prod_{a=1}^{N_f}
	\Gamma_h(\mu_a + \sigma_i)
	\Gamma_h(\nu_a - \sigma_i)
	\right)
\end{align}
On the magnetic side we find
\begin{align}
\label{ddd}
	Z_{\text{mag}} =&
	\prod_{a,b=1}^{N_f} \Gamma_h(\mu_a + \nu_b + j\omega\tau)
	\frac{\Gamma_h(\tau)^{\tilde{N}_c}}{\tilde{N}_c!}
	\int \prod_{i=1}^{\tilde{N}_c}	d\tilde{\sigma}_i
	\prod_{i<j}	\frac{\Gamma_h(\tau \pm \tilde{\sigma}_i-\tilde{\sigma}_j))}
	{\Gamma_h(\pm (\tilde{\sigma}_i-\tilde{\sigma}_j))}
\nonumber\\&
	\int d\xi 
	\prod_{j=0}^{n-1}
	\Gamma_h\left(
	\pm \xi
	-\frac{1}{2} \sum (\mu_a - \nu_a)
	+\tau (j-N_c+1)
	+\omega N_f
	\right)
\nonumber\\&
	\int dx
	e^{\pi i (2x \Lambda-\kappa x^2-2 \xi (x +  \sum_i \tilde{\sigma}_i))}
	\prod_{i=1}^{\tilde{N}_c} 
	\bigg(
	\prod_{a=1}^{N_f}
	\Gamma_h(\tau-\nu_a + \tilde{\sigma}_i)
	\Gamma_h(\tau-\mu_a - \tilde{\sigma}_i)
	\bigg)
\end{align}
The equivalence between (\ref{eee}) and (\ref{ddd}) 
 represents the equivalence of  the partition functions 
 of the electric and of the magnetic models obtained above in this section.

 \subsection{The case of $U(N_c)$ adjoint SQCD}	
	
We then consider $N_f+k_1$ flavors and assign a large positive (negative) mass to  $k_1$ of such fundamentals and anti-fundamentals. On the magnetic side $k_1$ dual quarks, acquire large negative   (positive)  real mass. Furthermore there are $k_1^2+2N_f k_1$ mesons with large positive (negative)  real mass.  The $T_j$ and $\tilde T_j$ fields have a large negative (positive) real mass as well. 
We end up with a duality between
\begin{itemize}
\item $U(N_c)_{\pm  k_1,\pm k_1+N_c k_2}$  SQCD with $N_f$ flavors $Q$ and $\tilde Q$, and adjoint $X$  and superpotential 
 $W = Tr X^{n+1}$
\item A dual gauge theory with gauge group 
\begin{equation}
	\overbracket{U(n(N_f+k_1)-N_c)_{\mp k_1}	 \times U}^{1}\! \overbracket{(1)_{n}
		\times 	U(1)_{k_2}}^{1}
\end{equation}
with $N_f$ dual flavors $q$ and $\tilde q$, and adjoint $Y$ of the non-abelian gauge group and $n N_f^2$ mesons 
$M_j = Q X^j \tilde Q$ with superpotential 
\begin{equation}
W = Tr Y^{n+1} + \sum_{j=0}^{n-1} M_j q Y^{j} \tilde q
\end{equation}
\end{itemize}
We have checked the validity of this duality by computing the real mass flow on the partition function.
In the following we restrict to the case of large positive real mass for the $k_1$ fundamentals, but the discussion
can be extended to the negative case straightforwardly.
First we have seen that the divergent terms in the large mass limit cancel between the electric and magnetic sides and 
then we have obtained an identity between  
\begin{align}
	Z_{\text{ele}}  =&
	\frac{\Gamma_h(\tau)^{N_c}}{N_c!}
	\int \prod_{i=1}^{N_c}	d\sigma_i
	e^{
		- \pi i  k_2
		\left( \sum_i\sigma_i \right)^2
		+
		2\pi i \Lambda
		\sum_i\sigma_i
		-
		\pi i k_1 \sum_i \sigma_i^2
	}
	\nonumber\\&
	\prod_{i<j}	\frac{\Gamma_h(\tau \pm (\sigma_i-\sigma_j))}
	{\Gamma_h(\pm (\sigma_i-\sigma_j))}
	\prod_{i=1}^{N_c} 
	\left(
	\prod_{a=1}^{N_f}
	\Gamma_h(\mu_a + \sigma_i)
	\Gamma_h(\nu_a - \sigma_i)
	\right)
\end{align}
and
\begin{align}
	Z_{\text{mag}} =&
	e^{-\frac{i\pi}{2} \phi}
	\prod_{a,b=1}^{N_f} \Gamma_h(\mu_a + \nu_b + j\omega\tau)
	\frac{\Gamma_h(\tau)^{\tilde{N}_c}}{\tilde{N}_c!}
	\int \prod_{i=1}^{\tilde{N}_c}	d\tilde{\sigma}_i
	\prod_{i<j}	\frac{\Gamma_h(\tau \pm \tilde{\sigma}_i-\tilde{\sigma}_j))}
	{\Gamma_h(\pm (\tilde{\sigma}_i-\tilde{\sigma}_j))}
	\nonumber\\&
	\int d\xi 
	\int dx
	e^{-2\pi i \xi x}
	e^{- 2  \pi i \xi \sum_i \tilde{\sigma}_i - \pi i n \xi^2}
	e^{\pi i k_2 x^2 +  2 \pi i x \Lambda}
	e^{- \pi i k_1 \sum_i \tilde{\sigma}_i^2}
	\nonumber\\&
	\prod_{i=1}^{\tilde{N}_c} 
	\left(
	\prod_{a=1}^{N_f}
	\Gamma_h(\tau-\nu_a + \tilde{\sigma}_i)
	\Gamma_h(\tau-\mu_a - \tilde{\sigma}_i)
	\right)
\end{align}
Where $\widetilde{N_c} = n(N_f+k_1) - N_c$ and
\begin{align}
	\phi =&
	-\frac{\pi i  n}{3 (n+1)^2}
	 ((n+1) (-3 k_1 (n+1) (\sum _{a=1}^{N_f} m_a^2+\sum _{a=1}^{N_f} n_a^2)
\nonumber\\&
	 +2 k_1 N_f (2 (4 n-1) \omega ^2-3 m_A ((n+1) m_A+2 (n-1) \omega ))
\nonumber\\&
	 +6 (n+1) N_f^2 (\omega -m_A)^2)-24 \omega  N_c ((n+1) N_f (\omega -m_A)+k_1 n \, \omega )
\nonumber\\&
	 +24 \omega ^2 N_c^2+\omega ^2 ((11 k_1^2+2) n^2+k_1^2-2))
\end{align}
%
%
%
%
%
%
\section{Chiral dualities}
\label{sec:chiral}
%
%
%
%
%
%

Another large class of models that can be investigated with the procedure explained in Section \ref{FTsec} 
consists  of $U(N_c)$ SQCD with a chiral matter content, i.e. with  a different number of fundamentals ($N_f$) and antifundamentals ($N_a$).
These models have been  investigated in \cite{Benini:2011mf} for SQCD and in \cite{Hwang:2015wna,Amariti:2020xqm} 
for adjoint SQCD.

For consistency with the literature we refer to such dualities by using the  notations of \cite{Benini:2011mf}
distinguishing three cases. From now on we consider only the case $k>0$, the case $k<0$ 
can be derived from this by applying parity and charge conjugation.
\begin{itemize}
\item $[p,q]$: in this case the electric theory has an $U(N_c)_k$ gauge group with $|N_f-N_a|<2k$ and $W=0$. The dual theory 
has gauge group $U\Big(\frac{N_f+N_a}{2}+k -N_c\Big)_{-k}$ and $W= M q \tilde q$.
\item $[p,0]$: in this case the electric theory has an $U(N_c)_{\frac{N_f-N_a}{2}}$ 
gauge group and $W=0$. The dual theory 
has gauge group $U\Big(N_f -N_c\Big)_{\frac{N_a-N_f}{2}}$ and $W= M q \tilde q + T t$.
\item $[p,q]^*$: 
 in this case the electric theory has an $U(N_c)_k$ 
gauge group with $|N_f-N_a|>2k$  and $W=0$. The dual theory 
has gauge group $U(\text{max}(N_f,N_a)-N_c)_{-k}$ and $W= M q \tilde q $.
\end{itemize}
The generalizations of these dualities in presence of adjoint matter have been denoted as 
$[p,q]_A$,$[p,0]_A$ and $[p,q]_A^*$ respectively in \cite{Amariti:2020xqm}.
In this case we have
\begin{itemize}
\item $[p,q]_A$: in this case the electric theory has an $U(N_c)_k$ gauge group with 
$|N_f-N_a|<2k$ and $W=Tr X^{n+1}$. The dual theory 
has gauge group $U\Big(n\Big(\frac{N_f+N_a}{2}+k\big) -N_c\Big)_{-k}$ and 
$W=Tr Y^{n+1}+ \sum_{j=0}^{n-1} M_j q Y^j \tilde q$.
\item $[p,0]_A$: in this case the electric theory has an $U(N_c)_{\frac{N_f-N_a}{2}}$ 
gauge group and $W=Tr X^{n+1}$. The dual theory 
has gauge group $U\Big(n N_f -N_c\Big)_{\frac{N_a-N_f}{2}}$ and 
$W= Tr Y^{n+1}+ \sum_{j=0}^{n-1} M_j q Y^j \tilde q+ T_j t_{n-1-j}$.
\item $[p,q]_A^*$: 
 in this case the electric theory has an $U(N_c)_k$ 
gauge group with $|N_f-N_a|>2k$  and $W=Tr X^{n+1}$. The dual theory 
has gauge group $U(n\,\text{max}(N_f,N_a)-N_c)_{-k}$ and 
$W= Tr Y^{n+1}+ \sum_{j=0}^{n-1} M_j q Y^j \tilde q+ T_j t_{n-1-j}$.
\end{itemize}
In the following we will just focus on the $[p,q]_A$, $[p,0]_A$ and $[p,q]_A^*$
dualities, observing that they reduce to the $[p,q]$, $[p,0]$ and $[p,q]^*$
by fixing $n=1$.

We proceed as follows. We first consider the non-chiral $U(N_c)$ duality
and gauge the topological $U(1)_J$ symmetry. Then we add a topological sector for the new topological $U(1)_{J'}$ symmetry
and after this step we gauge it. In this way we have a CS term and an FI term for the $U(1)_{J'}$ symmetry (in addition to the 
mixed CS between $U(1)_J$ and $U(1)_{J'}$). At this point, in the electric theory, we can integrate out the massive vector fields from  $U(1)_J$ and $U(1)_{J'}$ and then we consider the real mass flow to the chiral case.
Furthermore we must consider also non-trivial vacua for the abelian gauge symmetries in both the electric and the magnetic phase. This is because, in order to keep the duality, we need to assign a large vacuum expectation value to the scalars in the vector multiplets of $U(1)_{J}$  and $U(1)_{J'}$.
Only after the real mass flow we  
end up with the electric $U(N_c)_{k_1,k_1+N_c k_2}$ adjoint SQCD with $N_f$ fundamentals, $N_a$ antifundamentals and $W =Tr X^{n+1}$. Depending on the relative value of  $|N_f-N_a|$ and $2k_1$ we must consider different flows and we have different dualities, generalizing the $[p,q]_A$, the $[p,0]_A$ and the $[p,q]_A^*$ cases.
In the following we will study these three cases separately.

 %
 %
 %
\subsection{The $[p,q]_A$ case}
%
%
%

We start our analysis with the $[p,q]_A$ case. We assign a positive
large real mass to 
$N_f-N_f^{(1)}$ fundamentals and a   positive large real mass to 
$N_f-N_f^{(2)}$ antifundamentals.
We have to consider a nonzero vacuum for the scalars in the vector multiplet
for the non-abelian symmetry.
The electric theory has gauge group $U(N_c)_{k_1,k_1+N_c k_2}$
with $N_f^{(1)}$ fundamentals and 
$N_f^{(2)}$ antifundamentals.
The CS level $k_1$ generated by the real mass flow is $k_1 = N_f-\frac{1}{2}(N_f^{(1)}+N_f^{(2)} )$
and $ |N_f^{(1)}-N_f^{(2)}|  < 2k =2N_f- N_f^{(1)}-N_f^{(2)} $.

On the magnetic side we have to consider a nonzero vacuum for the scalars in the vector multiplets
for the non-abelian and for the abelian symmetries.
We are left with
 \begin{equation}
 \overbracket{U(k_1+\tfrac{1}{2}(N_f^{(1)}+N_f^{(2)})-N_c)_{-k_1} \times U}^{1}
 \overbracket{(1)_{-n}\times U(1)_{k_2}}^{1}
 \end{equation}
  gauge symmetry with mix CS levels as in the formula above.
There are $N_f^{(1)}$ dual antifundamentals and $N_f^{(2)}$ dual fundamentals and 
there is a superpotential $W = \beta Tr Y + Y^{n+1} + \sum_{j=0}^{n-1} M_j q Y^j \tilde q$, where $\beta$ is a singlet 
that is necessary to impose Tr$Y$ out of the chiral ring, as already discussed in \cite{Amariti:2020xqm}. 
One can then integrate out the massive singlets $\beta$ and $Tr Y$ on the dual partition function, and 
consider a traceless adjoint field $Y$.
There are also nontrivial contact terms
in the two-point functions of the global symmetry currents \cite{Closset:2012vp,Closset:2012vg}.

This real mass flow can be  concretely visualized  on the three-sphere partition function
by assigning the real masses as
\begin{equation}
\label{pqrm}
\left\{
\begin{array}{llll}
m_A & \to & m_A+ \frac{2N_f -N_f^{(1)}-N_f^{(2)}}{2 N_f} s;& \\
x & \to & x-\frac{N_c\Big(N_f^{(1)}-N_f^{(2)}\Big)}{2N_f}s; & \\
m_a & \to & m_a-\frac{N_f-N_f^{(1)}}{N_f} s,& a=1,\dots N_f^{(1)};  \\
m_a & \to & m_a+\frac{N_f^{(1)}}{N_f}s ,& a= N_f^{(1)}+1,\dots N_f;  \\
n_a & \to & n_a -\frac{N_f-N_f^{(2)}}{N_f} s,& a=1,\dots N_f^{(2)};  \\
n_a & \to & n_a+\frac{N_f^{(2)}}{N_f} s ,& a= N_f^{(2)}+1,\dots N_f ; \\
\sigma_i & \to & \sigma_i-\frac{N_f^{(1)}-N_f^{(2)}}{2 N_f }s;  & \\
\tilde{\sigma}_i & \to &\tilde \sigma_i-\frac{N_f^{(1)}-N_f^{(2)}}{2 N_f }s;  & \\
\xi & \to & \xi-\frac{N_f^{(1)}-N_f^{(2)}}{2} s.   \\
\Lambda & \to & \Lambda-\frac{(N_f^{(1)}-N_f^{(2)})(N_f + N_c k_2)}{2N_f} s.   \\
\end{array}
\right.
\end{equation}
By computing the large $s$ limit on the partition function we check that the divergent terms cancel between the electric and the magnetic phase.
We are left  with the identity between
\begin{align}
Z_\text{ele} =&\ 
\frac{\Gamma_h( \tau )^{N_c-1}}{N_c!}
\int \prod_{i=1}^{N_c} 
d\sigma_i \,  \prod_{i<j} \frac{\Gamma_h( \tau \pm  (\sigma_i -\sigma_j))}{\Gamma_h( \pm  (\sigma_i -\sigma_j))} 
\,e^{-\pi i k_1 \sum_i \sigma_i^2-\pi i k_2 (\sum_i \sigma_i)^2+\pi i \eta_1 \sum_i \sigma_i} \cdot   \nonumber \\
 & \cdot \prod_{i=1}^{N_c}\left( \prod_{a=1}^{N_f^{(1)}} \Gamma_h(\mu_a + \sigma_i)  
 \cdot
\prod_{b=1}^{N_f^{(2)}} \Gamma_h(\nu_a-\sigma_i)\right)\ ,
\end{align}
 with
 $\eta_1 = 
 \left(N_f^{(1)}-N_f^{(2)}\right) \left(m_A-\omega \right)+2 \Lambda$
 and
\begin{align}
Z_\text{mag} =&\ \frac{e^{\pi i n \phi} \Gamma_h( \tau )^{\widetilde N_c-1} }{ \widetilde N_c!}
\prod_{j=0}^{ n}  \prod_{a=1}^{N_f^{(1)}} \prod_{b=1}^{N_f^{(2)}} \Gamma_h(\mu_a+\nu_b+j \tau)
\int d x 
d\xi \left( \prod_{i=1}^{ \widetilde N_c}  d\tilde \sigma_i 
e^{-  \pi i ((2 \xi+\eta_2) \sigma _i-k_1 \sigma_i^2)}
\right)
\cdot 
\nonumber \\
&\cdot
e^{
- \pi i  k_2  x^2
+2  \pi i  \Lambda  x
-2  \pi i  \xi  x
+ \pi i  n \xi ^2
}
\nonumber \\
&\cdot
\prod_{i=1}^{ \widetilde N_c} 
\left(
\prod_{a=1}^{N_f^{(1)}} \Gamma_h(\tau-\mu_a - \tilde \sigma_i)
\cdot
\prod_{b=1}^{N_f^{(2)}} \Gamma_h(\tau-\nu_b+\tilde \sigma_i)
\right)
 \prod_{i<j} 
\frac{\Gamma_h( \tau \pm  (\tilde \sigma_i - \tilde \sigma_j))}{\Gamma_h( \pm  (\tilde \sigma_i - \tilde \sigma_j))} 
\ ,
\end{align}
with $\eta_2 =  \left(N_f^{(1)}-N_f^{(2)}\right) \pi (m_A-\tau +\omega)$.
The contribution of the global  CS are obtained from the exponential $e^{\pi i n \phi}$ 
and it reads
\begin{eqnarray}
\phi &=& 
\tau   \Bigg(
  \tau  N_c (N_c-n N_f)
-\tau\frac{N_f^{(1)}+ N_f^{(2)}}{8}     \left(4 N_c+(n-1)^2 N_f\right)
\nonumber \\
&+&\frac{\tau n^2   N_f^2}{2} +\frac{\omega (n-1)}{12}    (2-N_f^2+N_f^{(1)} N_f^{(2)})+\frac{1}{2}  N_f^{(1)} N_f^{(2)} \tau \Bigg)
\nonumber \\
&-&
\frac{   \omega  m_A ((N_f^{(1)}+N_f^{(2)}) ((n-1) N_f-2 N_c)+4 N_f^{(1)} N_f^{(2)})}{n+1}
\nonumber \\
&-&\frac{ \omega  m_A^2 }{2}  \left(\left(N_f^{(1)}+N_f^{(2)}\right) N_f-4 N_f^{(1)} N_f^{(2)}\right)
\nonumber \\
&-&\frac{N_f-N_f^{(2)}}{2}     \sum_{a=1}^{N_f^{(1)}} m_a^2
-
\frac{N_f-N_f^{(1)}}{2}  \sum _{b=1}^{N_f^{(2)}} n_b^2
\end{eqnarray}

 %
 %
 %
\subsection{The $[p,0]_A$ case}
%
%
%

In order to obtain the generalization of the  $[p,q]_A$ case we assign a  positive
 large real mass to $N_f-N_f^{(1)}$  antifundamentals.
We have to consider a nonzero vacuum for the scalars in the vector multiplet
for the non-abelian symmetry.
The electric theory has gauge group $U(N_c)_{k_1,k_1+N_c k_2}$
with $N_f$ fundamentals and 
$N_f^{(1)}$ antifundamentals.
The CS level $k_1$ generated by the real mass flow is $k_1  = \tfrac{1}{2}(N_f-N_f^{(1)} )$
and $ |N_f-N_f^{(1)}|  = 2k$.

On the magnetic side we have again to consider a nonzero vacuum for the scalars in the vector multiplets
for the non-abelian and for the abelian symmetries.
We are left with
\begin{equation}
\overbracket{U(n(k_1+\tfrac{1}{2}(N_f+N_f^{(1)}))-N_c)_{-k_1} \times U}^{1}
\overbracket{(1)_{-\frac{n}{2} }\times U(1)_{k_2 }}^{1}
\end{equation}
 gauge symmetry with mix CS levels as in the formula above.
 In the non-abelian sector there are $N_f$ dual antifundamentals and $N_f^{(1)}$ dual fundamentals, and a traceless adjoint
 $Y$.
  There is also a set of fields charged under $U(1)_{-\frac{n}{2}}$, denoted as $t_{j}$, interacting with a set of singlets $T_j$, corresponding to the monopoles of the electric theory.
 The dual superpotential in this case is $W = \text{Tr} Y^{n+1} + \sum_{j=0}^{n-1} M_j q Y^j \tilde q + T_{j} t_{n-1-j}$.

The real mass flow just discussed can be  concretely visualized  on the three-sphere partition function
by assigning the real masses as
\begin{equation}
\label{p0rm}
\left\{
\begin{array}{llll}
m_A & \to &m_A+\frac{N_f-N_f^{(1)}}{2N_f}s;& \\
x & \to &x+\frac{N_c(N_f-N_f^{(1)})}{2N_f} s; & \\
n_a  & \to &n_a -\frac{N_f-N_f^{(1)}}{N_f}s,& a=1,\dots N_f^{(1)};  \\
n_a & \to &n_a+\frac{N_f^{(1)}}{N_f}s, & a=N_f^{(1)}+1,\dots,N_f-N_f^{(1)};\\
{\sigma}_i & \to & \sigma_i- \frac{N_f-N_f^{(1)}}{2N_f} s;  & \\
\tilde{\sigma}_i & \to &\tilde \sigma_i- \frac{N_f-N_f^{(1)}}{2N_f} s;  & \\
\xi & \to & \xi+\frac{N_f-N_f^{(1)}}{2}s.& \\
\Lambda & \to & \Lambda+\frac{(N_f-N_c k_2)(N_f-N_f^{(1)})}{2 N_f}s.& \\
\end{array}
\right.
\end{equation}

\begin{align}
Z_\text{ele} =& \ 
\frac{\Gamma_h( \tau)^{N_c-1}}{N_c!}
\int \prod_{i=1}^{N_c} d\sigma_i \,
\prod_{i<j} \frac{\Gamma_h( \tau\pm  (\sigma_i -\sigma_j))}{\Gamma_h( \pm  (\sigma_i -\sigma_j))}\, e^{-\pi i (k_1 \sum_i \sigma_i^2+k_2 (\sum_i \sigma_i)^2- \eta_1 \sum_i \sigma_i )} \cdot  \nonumber \\
 &\cdot 
 \prod_{i=1}^{N_c} \left(
 \prod_{a=1}^{N_f} \Gamma_h(\mu_a + \sigma_i)  
 \cdot
\prod_{b=1}^{N_f^{(1)}} \Gamma_h(\nu_a-\sigma_i)\right) \ ,
\end{align}
with $\eta_1 = 2 \Lambda +(N_f^{(1)}-N_f)(m_A-\omega)$
\begin{align}
Z_\text{mag} =&\ \frac{e^{\pi i n \phi}\Gamma_h( \tau)^{\widetilde N_c-1}}{ \widetilde N_c!}
\prod_{j=0}^{n-1}  \prod_{a=1}^{N_f} \prod_{b=1}^{N_f^{(1)}} \Gamma_h(\mu_a+\nu_b+j \tau) \cdot 
\int  
  dx  d\xi  
\prod_{i=1}^{  \widetilde N_c} \Big(
d\tilde \sigma_i 
e^{\pi i (k_1  \tilde \sigma_i^2+(\eta_2-2\xi) \tilde \sigma_i)}
\cdot
\nonumber \\ & 
\cdot
\prod_{i<j} \frac{\Gamma_h( \tau\pm  (\tilde \sigma_i - \tilde \sigma_j))}
{\Gamma_h( \pm  (\tilde \sigma_i -\tilde \sigma_j))}
\prod_{a=1}^{N_f} \Gamma_h(\tau-\mu_a- \tilde \sigma_i)
\cdot
\prod_{b=1}^{N_f^{(1)}} \Gamma_h(\tau-\nu_b+\tilde \sigma_i)
\Big)
\cdot
 \\ & 
\cdot
  e^{-\pi i  k_2 x^2+\frac{1}{2}  \pi i  n \xi ^2+2 \pi i  \Lambda  x-2  \pi i  \xi  x} 
\prod _{j=0}^{n-1} \Gamma _h \left(N_f \left(\omega -m_A\right)+\tau  \left(-N_c+j+1\right)+\xi \right)
\nonumber
\end{align}
with $\eta _2=(N_f-N_f^{(1)}) (\omega -\tau+m_A)$.
The contribution of the global  CS is obtained from the exponential $e^{ \pi i n \phi}$ 
and it reads
\begin{eqnarray}
\phi &=&
\frac{   \omega ^2 (
12 N_c(N_c-n N_f)-2 N_f^{(1)} (6 N_c+(n^2-3 n-4) N_f)
+(n^2-1) (5 N_f^2+1))}{6 (n+1)^2}
\nonumber  \\
&-&\frac{    \omega  m_A  (N_f^{(1)}   ((n+3) N_f\!-\!2 N_c )\!-\!2 N_f^2 )}{n+1}
+
\frac{ m_A^2 (3 N_f^{(1)} \!\!-\!2 N_f) N_f}{2}   
\!-\!\frac{ (N_f-N_f^{(1)} )}{2}    \sum _{b=1}^{N_f} n_b^2
\nonumber\\
\end{eqnarray}

 %
 %
 %
\subsection{The $[p,q]_A^*$ case}
%
%
%

In order to obtain the generalization of the  $[p,q]_A^*$ case we assign a positive
 large real mass to $N_f^{(1)}$ antifundamentals and a negative large real mass to 
$N_f^{(2)}$ antifundamentals.

We have to consider a nonzero vacuum for the scalars in the vector multiplet
for the non-abelian symmetry.
The electric theory has gauge group $U(N_c)_{k_1,k_1+N_c k_2}$
with $N_f$ fundamentals and 
$N_a$ antifundamentals.
The CS level $k_1$ generated by the real mass flow is $k_1  = \frac{1}{2}(N_f^{(1)}-N_f^{(2)} )$
and we have$N_f$ fundamentals, $N_a = N_f-N_f^{(1)}-N_f^{(2)}$ antifundamentals, with   $ N_f-N_a  >k_1$.

On the magnetic side we have to consider a nonzero vacuum for the scalars in the vector multiplets
for the non-abelian and for the abelian symmetries.
We are left with
\begin{equation}
\overbracket{U(n N_f-N_c)_{-k_1} \times U}^{1}
\overbracket{(1)_{0}\times U(1)_{k_2}}^{1}
\end{equation}
 gauge symmetry with mixed CS levels as in the formula above.
There are $N_f$ dual antifundamentals and $N_a$ dual fundamentals and 
there is a superpotential $W = Y^{n+1} + \sum_{j=0}^{n-1} M_j q Y^j \tilde q$, again with a traceless adjoint $Y$. 
Observe that the duality for the case with  $ N_f<N_a$ can be obtained by acting with parity, such that the non-abelian dual gauge group becomes $U(n \, \text{max}(N_f , N_a)-N_c)_{-k_1}$.

This real mass flow can be  concretely visualized  on the three-sphere partition function
by assigning the real masses as

\begin{equation}
\label{pq*rm}
\left\{
\begin{array}{llll}
m_A & \to & m_A+\frac{N_f^{(1)}-N_f^{(2)}}{2N_f}s;& \\
x & \to &x-\frac{N_c(N_f^{(1)}-N_f^{(2)})}{2N_f}s; & \\
n_a & \to &n_a - \frac{N_f^{(1)}-N_f^{(2)}}{N_f}s,& a=1,\dots, N_f-N_f^{(1)}-N_f^{(2)} =N_a; \\
n_a & \to &n_a+\frac{N_f-N_f^{(1)}+N_f^{(2)} }{N_f} s,& a=N_a+1,\dots,N_f^{(1)}+N_a = N_f-N_f^{(2)};\\
n_a & \to &n_a-\frac{N_f+N_f^{(1)}-N_f^{(2)} }{N_f} s,& a= N_f-N_f^{(2)}+1,\dots,N_f;\\
{\sigma}_i & \to & \sigma_i-\frac{N_f^{(1)}-N_f^{(2)}}{2N_f}s;  & \\
\tilde{\sigma}_i & \to &\tilde \sigma_i-\frac{N_f^{(1)}-N_f^{(2)}}{2N_f}s;  & \\
\xi & \to & \xi-\frac{N_f^{(1)}+N_f^{(2)}}{2}.& \\
\Lambda &\to& \Lambda -\frac{s \left(N_f \left(N_f^{(1)}+N_f^{(2)}\right)+\left(N_f^{(1)}-N_f^{(2)}\right)
N_c k_2\right)}{2 N_f}\\
\end{array}
\right.
\end{equation}

\begin{align}
Z_\text{ele} =&\ 
\frac{\Gamma_h( \tau)^{N_c-1}}{N_c!}
\int \prod_{i=1}^{N_c} d\sigma_i \prod_{i<j} 
\frac{
\Gamma_h( \tau\pm  (\sigma_i -\sigma_j))
}
{
\Gamma_h( \pm  (\sigma_i -\sigma_j))
}
\,
e^{-\pi i (  \sum_i \sigma_i (k_1 \sigma_i-2 \eta_1) + k_2 (\sum_i \sigma_i)^2)} \cdot \nonumber \\
&\prod_{i=1}^{N_c}\left(
  \prod_{a=1}^{N_f} \Gamma_h(\mu_a + \sigma_i)  
 \cdot
\prod_{b=1}^{N_a} \Gamma_h(\nu_a-\sigma_i)
\right)\ ,
\end{align}
with $\eta_1 =  \Lambda+ k_1 \left(m_A-\omega \right)$
and
\begin{align}
Z_\text{mag} =&\ \frac{e^{\pi i n k_1   \phi} \Gamma_h( \tau)^{\widetilde N_c-1}}{\widetilde N_c!}
\prod_{j=0}^{n-1}  \prod_{a=1}^{N_f} \prod_{b=1}^{N_a} 
\Gamma_h(\mu_a+\nu_b+j  \tau) \cdot
\int dx d\xi 
e^{-\pi i  k_2 x^2+2 \pi i  \Lambda  x-2 \pi i  \xi  x}
\nonumber
\\
&\cdot
\Bigg( \prod_{i=1}^{\widetilde N_c} d\tilde \sigma_i \prod_{i<j} 
\,
e^{\pi i  \sum_i \tilde \sigma_i (k_1 \tilde \sigma_i-2 (\xi + \eta_2) ) }\cdot 
 \prod_{a=1}^{N_f} \Gamma_h(\tau-\mu_a-\tilde \sigma_i)  
 \cdot
\prod_{b=1}^{N_a} \Gamma_h(\tau-\nu_a + \tilde \sigma_i)
\Bigg) 
\nonumber
\\
&\cdot
 \frac{
\Gamma_h( \tau\pm  (\tilde \sigma_i -\tilde \sigma_j))
}
{
\Gamma_h( \pm  (\tilde \sigma_i -\tilde \sigma_j))
}
\ ,
\end{align}
with $\eta_2 = (m_A- \tau + \omega)$.
The contribution of the global  CS are obtained from the exponential $e^{\pi i n k_1 \phi}$ 
and it reads
\begin{eqnarray}
\phi=
\tau^2 N_c+ \omega \tau \frac{  N_f (n-4) }{3}
+ \tau  m_A ((n+3) N_f-2 N_c)
-3 n m_A^2 N_f
-\sum _{a=1}^{N_f} m_a^2
\end{eqnarray}
%
%
%
%
%
%
\section{Exceptional dualities}
\label{sec:ex}
%
%
%
%
%
%
In this section we consider the  dualities studied in \cite{Amariti:2018wht,Benvenuti:2018bav}
and extend our analysis to one of them, considering $U(N_c)_0$  with 2 fundamentals, 2 antifundamentals and an adjoint (the case without adjoint was discussed originally in 
\cite{Dimofte:2012pd}). This is not a special case of the theories considered in section \ref{sec:adj} because here the superpotential vanishes.
This model  is involved in a triality with other two gauge theories. Such a 
triality originates from the \emph{parent}  duality of $USp(2N_c)$ with 8 fundamentals, one antisymmetric and $W=0$ and a theory with the same gauge group and charged matter with additional singlets that couple to the dual fundamentals via a Seiberg-like superpotential. 
If $N_c$ is even\footnote{In this section we will always assume that $N_c$ is even when considering self-dualities.}  this duality can be brought to a self-dual form by flipping half of the singlets of the dual theory, therefore the global symmetry of $USp(N_c)$ with 8 fundamentals and an adjoint enhances to $E_7 \times U(1)$.
 Performing real mass flows generates a web of theories  
 classified in \cite{Amariti:2018wht} by  following the notation of \cite{VanDeBult} in terms of 
 the breakdown of $E_7$ to sub-algebras. 
 Two models  with the same sub-algebra are believed to be dual and furthermore they can enjoy global symmetry enhancement (which again signals the existence of self-dualities for the theories themselves). 

Here we  consider the triality found for the $A_3 \times A_1$ case, that relates:
\begin{itemize}
	\item $U(N_c)_0$ gauge theory with two fundamentals $Q_{1,2}$, two anti-fundamentals  
	$\tilde{Q}_{1,2}$, one adjoint $X$ and $W=0$
	\item $U(N_c)_{\frac{1}{2}}$ gauge theory with 
	three fundamentals $p_{0,1,2}$, two anti-fundamentals  $\tilde{p}_{1,2}$, one adjoint $Z$ and  \begin{equation}
	W = \tilde{t}
	+ \sum_{j=0}^{N_c-1}	\sum_{i=0}^{1} M^j_{i1} p_i Z^j \tilde{p}_1
	+ \sum_{j=0}^{N_c-1}	\hat{T}^j p_2 Z^j \tilde{p}_0
\end{equation}
	\item $USp(2N_c)_2$ gauge theory with four fundamentals $q_{1,2,3,4}$, one antisymmetric $Y$ and 
\begin{equation}	
W= \sum_{j=0}^{N_c-1} 	M_j q_1 Y^{N_c-1-j} q_2 + N_j q_3 Y^{N_c-1-j} q_4 
\end{equation}
\end{itemize}
where $M_{ij}$ and $\hat{T}^j$ are singlets in the $U(N_c)_{\frac{1}{2}}$ theory, $\tilde{t}$ is the anti-monopole of the $U(N_c)_{\frac{1}{2}}$ theory and $M_j, N_j$ are singlets in the $USp(2N_c)_2$ theory. The global symmetry of all these theories enhances to $A_3 \times A_1$, therefore they will enjoy self-duality relations. For example the $USp(2N_c)_2$ theory enjoys a $A_3 \rightarrow A_3\times A_1$ global symmetry enhancement, where the generator of the additional $A_1$ corresponds to the duality between:
\begin{itemize}
	\item $USp(2N_c)_2$ gauge theory with four fundamentals $q_{1,2,3,4}$, one antisymmetric and $W=0$
	\item $USp(2N_c)_2$ gauge theory with four fundamentals $\tilde{q}_{1,2,3,4}$, one antisymmetric $Y$, singlets $\tilde{M}^j_{il}$ and $$W= \sum_{j=0}^{N_c-1} \sum_{i<l} \tilde{M}^j_{il} \tilde{q}_i \tilde{Y}^{N_c-1-j} \tilde{q}_l$$
\end{itemize}

Similarly to the $USp(2N_c)_0$ theory with 8 fundamentals discussed above, for $N_c$ even one can flip half of the mesons, say the last $N_c/2$ ones, and bring this theory to a self-dual form:
\begin{equation}	\label{eq:selfdual_USp}
	\begin{gathered}
		USp(2N_c)_2 \\
		\text{four fundamentals } q_{1,2,3,4},
		\text{ 1 antisymmetric } Y \\
		\text{singlets } \tilde{M}^j_{il}, \quad j=0,\dots, \frac{N_c}{2}-1 \\
		W= \sum_{j=0}^{\frac{N_c}{2}-1} \sum_{i<l} M^j_{il} q_i Y^{N_c-1-j} q_l
	\end{gathered}
\end{equation}

This deformation can be mapped to the $U(N_c)_0$ and to the $U(N_c)_\frac{1}{2}$ theories via the dictionary:
\begin{align}
	\begin{gathered}
	U(N_c)_0\\
	\left\{
		\begin{array}{c}
			T_j=T X^{N_c-1-j} \\
			\tilde{T}_j=\tilde{T} X^{N_c-1-j} \\
			Q_i X^j \tilde{Q}_1 \\
			Q_i X^j \tilde{Q}_2 \\
			s^j_{i,1}\\
			s^j_{i,2}
		\end{array}
	\right\}
	\end{gathered}
	\quad \longleftrightarrow \quad
	\begin{gathered}
	USp(2N_c)_2\\
	\left\{
	\begin{array}{c}
		M_j\\
		N_j\\
		S^j_{i1}\\
		S^j_{i2}\\
		q_i Y^j q_3\\
		q_i Y^j q_4
	\end{array}
	\right\}
	\end{gathered}
	\quad \longleftrightarrow \quad
	\begin{gathered}
	U(N_c)_{\frac{1}{2}}\\
	\left\{
	\begin{array}{c}
		\hat{T}_j\\
		p_2 Z^j \tilde{p}_1\\
		M^j_{i-1,1}\\
		p_{i-1}Z^j \tilde{p}_0\\
		p_{i-1} Z^j \tilde{p}_1\\
		\hat{s}^j_{i0}
	\end{array}
	\right\}
	\end{gathered}
\end{align}
The resulting theories will be dual to the $USp$ theory in its self-dual form, so they will be self-dual themselves. They are:
\begin{equation}\label{eq:selfdual_U0}
	\begin{gathered}
		U(N_c)_0 \\
		\text{two fundamentals } Q_{1,2} \text{ and two anti-fundamentals } \tilde Q_{1,2}\\
		\text{1 adjoint }X\\
		W= \sum_{j=\frac{N_c}{2}}^{N_c-1} T_j m_j + \tilde{T}_j n_j
		+ \sum_{j=0}^{\frac{N_c}{2}-1} \sum_{i,l=1}^{2} s_{il}^j Q_i X^j \tilde{Q}_l
	\end{gathered}
\end{equation}

and:
\begin{equation}\label{eq:selfdual_U12}
	\begin{gathered}
		U(N_c)_{\frac{1}{2}}\\
		\text{three fundamentals }  p_{0,1,2}   \text{ and two anti-fundamentals } \tilde p_{1,2}\\
		\text{1 adjoint } Z \\
		W = \tilde{t}
		+ \sum_{j=\frac{N_c}{2}}^{N_c-1}	\sum_{i=0}^{1} M^j_{i1} p_i Z^j \tilde{p}_1
		+ \sum_{j=\frac{N_c}{2}}^{N_c-1}	\hat{T}^j p_2 Z^j \tilde{p}_0
		+ \sum_{j=\frac{N_c}{2}}^{N_c-1}	n_j p_2 Z^j \tilde{p}_1
		+ \sum_{j=0}^{\frac{N_c}{2}-1} 		\sum_{i=0}^{1} 	\hat{s}^j_{i+1,0} p_i Z^j \tilde{p}_0
	\end{gathered}
\end{equation}

\begin{table}[!ht] 
	\begin{center}
	\addtolength{\tabcolsep}{20pt}
	\setlength\extrarowheight{5pt}
	\addtolength{\leftskip} {-2cm} 
	\addtolength{\rightskip}{-2cm}
		\begin{tabular}{c||cc|c}
			\textbf{Theory} & \multicolumn{2}{c|}{\textbf{Explicit Symmetries}} & \textbf{Degeneracy}
			\\
			\hline
			IR f.p. & $A_3$ & $A_1$ & 1
			\\
			\hline
			\multirow{2}{*}{$USp(2N_c)_0$} & $A_3$ & self-duality & \multirow{2}{*}{2}
			\\
			& ($\tilde{\mu}_1,\tilde{\mu}_2,\tilde{\mu}_3,\tilde{\mu}_4$) & \eqref{eq:selfdual_USp}  &
			\\
			\hline
			\multirow{2}{*}{$U(N_c)_0$} & $A_1 \times A_1$ & self-duality & \multirow{2}{*}{12}
			\\
			& $(\mu_1, \mu_2)\times (\nu_1,\nu_2)$ & \eqref{eq:selfdual_U0}  &\\
			\hline
			\multirow{2}{*}{$U(N_c)_\frac{1}{2}$} & $A_2$ & $A_1$ & \multirow{2}{*}{4}
			\\
			& $(\hat{\mu}_0,\hat{\mu}_1,\hat{\mu}_2)$ & $(\hat{\nu}_0,\hat{\nu}_1)$  &
		\end{tabular}
	\end{center}
\caption{non-abelian global symmetry in the triality.}
\label{tab:triality_sym_map}
\end{table}

The non-abelian symmetry maps across the triality together with the degeneracy of the theories is listed in Table \ref{tab:triality_sym_map}.
We can perform our procedure in the $U(N_c)_0$ theory because it does not contain a monopole superpotential  and hence it does not break the topological symmetry. 

Observe that models with higher degree of symmetry  in the classification scheme discussed in 
 \cite{Amariti:2018wht,Benvenuti:2018bav} all have a monopole superpotential which obstructs the gauging of the topological $U(1)_J$ 
 while theories with a lower degree of symmetry can be obtained from the discussion here by an appropriate real mass flows.
 
  In order to obtain the corresponding triality with a non-zero CS term for the diagonal $U(1)$ in the $U(N_c)_0$ theory we gauge the topological symmetry $U(1)_J$. We then add  a CS contact term for the new topological symmetry $U(1)_{J'}$ and finally gauge this symmetry as well. The resulting model  has gauge group $U(N_c)_{0,N_c k_2}$, while the dual phases have two additional $U(1)$ gauge sectors. We end up with a triality between:
\begin{itemize}
	\item $U(N_c)_{0,N_c k_2}$ gauge theory with two fundamentals $Q_{1,2}$, two anti-fundamentals  
	$\tilde{Q}_{1,2}$, one adjoint $X$ and $W=0$
	\item $\overbracket{U(N_c)_{\frac{1}{2}} \times U(1)_{-22 N_c}   }^{-5} \overbracket{\times U(1)_{k_2}}^{1}$ gauge theory with 
	 three fundamentals $p_{0,1,2}$, two anti-fundamentals  
	$\tilde{p}_{1,2}$, one adjoint $Z$ and $$W = \tilde{t}
	+ \sum_{j=0}^{N_c-1}	\sum_{i=0}^{1} M^j_{i1} p_i Z^j \tilde{p}_1
	+ \sum_{j=0}^{N_c-1}	\hat{T}^j p_2 Z^j \tilde{p}_0$$
	(if $k_2 = 144 \hat{k}_2$ with $\hat{k}_2 \in \mathbb{Z}$)
	\item 
	$USp(2N_c)_2\times \overbracket{ U(1)_{-2 N_c} \times U(1)_{k_2}}^{1}$ gauge theory with four fundamentals $q_{1,2,3,4}$, one antisymmetric $Y$ and $$	W= \sum_{j=0}^{N_c-1} 	M_j q_1 Y^{N_c-1-j} q_2 +
	N_j q_3 Y^{N_c-1-j} q_4 $$
	(if $k_2 = 4 \hat{k}_2$ with $\hat{k}_2 \in \mathbb{Z}$)
\end{itemize}
Notice that the $U(N_c)_{\frac{1}{2}}$ theory contains matter that is charged under the topological $U(1)_J$ that we gauged, therefore the additional $U(1)\times U(1)$ gauge sector can't be integrated out as we did in the $U(N_c)_0$ theory. Moreover we cannot bring the $U(N)_{0,k_2}$ theory to a self-dual form because the deformation required to do so contains the $U(N_c)_0$ monopoles which are charged under the $U(1)_J$. The self-duality then becomes a regular duality between:
\begin{itemize}
	\item $U(N_c)_{0,N_c k_2}$ gauge theory with (2,2) flavors, one adjoint  and $W=0$
	\item $\overbracket{U(N_c)_{0} \times U(1)_0   }^{1} \overbracket{\times U(1)_{k_2}}^{1}$ gauge theory with (2,2) flavors $Q_i$, $\tilde{Q}_i$, one adjoint $X$ and $$W= \sum_{j=0}^{N_c-1} T_j m_j + \tilde{T}_j n_j
	+ \sum_{j=0}^{N_c-1} \sum_{i,l=1}^{2} s_{il}^j Q_i X^j \tilde{Q}_l$$
\end{itemize}
We can analyze these exceptional dualities at the level of the partition function. The self-duality for the $USp$ theory with 4 fundamentals is given by \textbf{Theorem 5.6.13} of \cite{VanDeBult}:
\begin{align}
\label{USPPF}
	Z_{USp(2N_c)_2} (\mu,\tau) =& \,Z_{USp(2N_c)_2} (\tilde{\mu},\tau)
	\prod_{j=0}^{N_c-1} \prod_{1\leq r<s\leq 4}
	\Gamma_h (j\tau + \mu_r + \mu_s) \nonumber \\
	\times &\, c(-2N_c \xi (2\omega + (N_c-1)\tau))	
\end{align}
where
\begin{align}
	\label{eq:selfdual_USp_reparam_1}
	\tilde{\mu} =& (\mu_1 + \xi,\mu_2 + \xi,\mu_3 + \xi,\mu_4 + \xi), 
	\\
	2\xi =& 2\omega - (N_c-1)\tau - \sum_{r=1}^{4} \mu_r 
	=  -2\omega + (N_c-1)\tau + \sum_{r=1}^{4} \tilde{\mu}_r
	\label{eq:selfdual_USp_reparam_2}
\end{align}
and $c(x) \equiv e^{-\frac{i \pi x}{2}}$.
In (\ref{USPPF}) we have referred to the partition function of an $USp(2N_c)$ gauge theory 
with vanishing CS level, four fundamentals with real masses $\mu_a$ and an antisymmetric with real mass $\tau$, denoting it by $Z_{USp(2N_c)_0} (\mu,\tau) $.
In general the partition function for an $USp(2N_c)_{2 \kappa}$ gauge theory with $2N_f$ fundamentals and an antisymmetric is given by
\begin{eqnarray}
Z_{USp(2N_c)_{2\kappa}} (\mu,\tau)
\!=\!
\frac{1}{2^{N_c} N_c!} \! \int \prod_{I=1}^{N_c}  
 \frac{d\sigma_i 
e^{-2 \pi i \kappa \sigma_i^2}}{\Gamma_h(\pm 2 \sigma_i)}
 \prod_{a=1}^{2 N_f} \Gamma_h(\pm \sigma_i + \mu_a) 
\!
\prod_{i<j} \frac{\Gamma_h(\pm \sigma_i \pm \sigma_j + \tau)}{\Gamma_h(\pm \sigma_i \pm \sigma_j)}
\nonumber \\
\end{eqnarray}
The mass parameter map is explicitly:
\begin{align}
	\tilde{\mu}_1 =& \frac{1}{2} \left(\phantom{-}\mu_1-\mu_2-\mu_3-\mu_4-\tau  \left(N_c-1\right) \right)+\omega 
	\nonumber\\
	\tilde{\mu}_2 =& \frac{1}{2} \left(-\mu_1+\mu_2-\mu_3-\mu_4-\tau  \left(N_c-1\right)\right)+\omega 
	\nonumber\\
	\tilde{\mu}_3 =& \frac{1}{2} \left(-\mu_1-\mu_2+\mu_3-\mu_4-\tau  \left(N_c-1\right)\right)+\omega 
	\nonumber\\
	\tilde{\mu}_4 =& \frac{1}{2} \left(-\mu_1-\mu_2-\mu_3+\mu_4-\tau  \left(N_c-1\right)\right)+\omega 
\end{align}
Flipping half of the singlets corresponds to multiplying both sides by:
\begin{equation}
	\prod_{j=\frac{N_c}{2}}^{N_c-1} \prod_{1\leq r<s\leq 4} \Gamma_h (2\omega - (j\tau + \mu_r + \mu_s))
\end{equation}
On the RHS some of the Hyperbolic Gamma functions simplify, what remains is:
\begin{equation}
	\prod_{j=0}^{\frac{N_c}{2}-1} \prod_{1\leq r<s\leq 4}
	\Gamma_h (j\tau + \mu_r + \mu_s) 
	=
	\prod_{j=0}^{\frac{N_c}{2}-1} \prod_{1\leq r<s\leq 4}
	\Gamma_h (2\omega - ((N_c - 1 - j)\tau + \tilde{\mu}_r + \tilde{\mu}_s))
\end{equation}
This has the same form of the term we originally multiplied by, but with $\mu$ exchanged for $\tilde{\mu}$. Finally we can rewrite the contact terms using $4\xi = \sum_r \tilde{\mu}_r - \sum_r \mu_r$ and obtain the identity:
\begin{align}	\label{eq:selfudal_USp_Z}
	Z_{USp(2N_c)_2} (\mu,\tau)&
	\prod_{j=\frac{N_c}{2}}^{N_c-1} \prod_{1\leq r<s\leq 4}
	\Gamma_h (2\omega - (j\tau + \mu_r + \mu_s)) 
	\nonumber\\
	&c(-\frac{N_c}{2} \sum_r \mu_r (2\omega + (N_c-1)\tau)) =
	\nonumber\\
	Z_{USp(2N_c)_2} (\tilde{\mu},\tau)&
	\prod_{j=\frac{N_c}{2}}^{N_c-1} \prod_{1\leq r<s\leq 4}
	\Gamma_h (2\omega - (j\tau + \tilde{\mu}_r + \tilde{\mu}_s)) 
	\nonumber\\
	&c(-\frac{N_c}{2} \sum_r \tilde{\mu}_r (2\omega + (N_c-1)\tau))
\end{align}

This identity represents an invariance of the integral under the re-parametrization given by \eqref{eq:selfdual_USp_reparam_1}-\eqref{eq:selfdual_USp_reparam_2} and corresponds to the self-duality property of the $USp$ theory \eqref{eq:selfdual_USp}. The presence of the contact term implies that the theory is self-dual only if it includes the correct global CS terms needed to produce such contact terms.

The identities between the partition function of the symplectic gauge theory and those of the two unitary theories are given by \textbf{Theorem 5.6.17} of \cite{VanDeBult}.
The first relation is
\begin{align} 	\label{eq:exceptional_Z_U0_USp_VandeBult}
	Z_{U\left(N_{c}\right)_{0}}(\mu, \nu ; \tau ; \lambda)=&
	Z_{USp\left(2 N_{c}\right)_{2}}\left(\mu_{\sigma^{\prime}} ; \tau\right)
	\nonumber\\&
	 \prod_{j=0}^{N_{c}-1} \Gamma_{h}\left(2 \omega \pm \frac{1}{2} \lambda-\frac{1}{2} \sum_{\alpha=1,2}\left(\mu_{\alpha}+\nu_{\alpha}\right)-\tau\left(N_{c}-1-j\right)\right) 
	e^{\frac{i\pi}{2} \phi_1}
\end{align}
where the partition function for the unitary gauge group has been defined in formula (\ref{zuad}).
The real masses are parameterized on the RHS of (\ref{eq:exceptional_Z_U0_USp_VandeBult}) as
\begin{align}
	4 \sigma' =& \nu_1 + \nu_2 - \mu_1 -\mu_2 - \lambda 
	\\
	\mu_{\sigma'} =& (\mu_1 + \sigma',\mu_2 + \sigma',\nu_1 - \sigma',\nu_2 - \sigma')
	\\
	\phi_1=&N_{c}\left(4 \sigma^{\prime 2}-2 \mu_{1} \mu_{2}-2 \nu_{1} \nu_{2}-\left(N_{c}-1\right) \tau 4 m_A-\frac{2}{3}\left(N_{c}-1\right)\left(N_{c}-2\right) \tau^{2}\right)
	\nonumber\\
	=& \frac{N_c}{4} \lambda^2 + \mathcal{O} (\lambda^0)
	\label{eq:exceptional_U0_USp_CT_lambda}
\end{align}
The second relation is
\begin{align}	\label{eq:exceptional_Z_U12_USp_VandeBult}
	Z_{U(N_c)_{1/2}} (\hat{\mu}, \hat{\nu}; \tau;\hat{\lambda} )
	=&
	Z_{USp(2N_c)_2}		(\hat{\mu}_\sigma;\tau)
	\prod_{j=0}^{N_c-1}	\prod_{r=0}^{2}
	\Gamma_h (j \tau + \hat{\mu}_r + \hat{\nu}_1)
	\nonumber\\
	&\times
	c(N_c(
	\hat{\lambda}^2 - (\hat{\lambda} + \hat{\nu}_0)	((N_c-1)\tau + 2\omega)
	+ \omega^2 + \frac{1}{2}(N_c - 1)\tau^2
	))
\end{align}
With:
\begin{align}
	2\sigma =& 2\omega - (N_c-1)\tau - \hat{\mu}_0- \hat{\mu}_1- \hat{\mu}_2- \hat{\nu}_1
	\\
	\hat{\mu}_\sigma =& (\hat{\mu}_0 + \sigma,	\hat{\mu}_1 + \sigma,
	\hat{\mu}_2 + \sigma,	\hat{\nu}_0 - \sigma)
\end{align}
Where the mass parameters and the FI term satisfy the balancing condition:
\begin{equation}	\label{eq:UN_UNhalf_bc}
	2(N_c-1)\tau + \sum_{r=0}^{2} \hat{\mu}_r + \sum_{s=0}^{1} \hat{\nu}_s
	= 
	\hat{\lambda} + (2 + 2 - 3) \omega
\end{equation}

By combining these identity it is straightforward to find the self-dualities for the $U(N_c)_{\frac{1}{2}}$ and the $U(N_c)_{0}$ theories, and also the duality between the two theories with unitary gauge groups, which reads:

\begin{align} 	\label{eq:UN0_UN1/2_raw}
	Z_{U\left(N_{c}\right)_{0}}(\mu , \nu ; \tau ; \lambda)=&
	Z_{U(N_c)_{1/2}} (\hat{\mu}, \hat{\nu}; \tau,\hat{\lambda})
	\prod_{j=0}^{N_c-1}	\prod_{r=0}^{2}
	\Gamma_h (2\omega - (j \tau + \hat{\mu}_r + \hat{\nu}_1))
	\nonumber\\&
	\prod_{j=0}^{N_{c}-1} \Gamma_{h}\left(2 \omega \pm \frac{1}{2} \lambda-\frac{1}{2} \sum_{\alpha=1,2}\left(\mu_{\alpha}+\nu_{\alpha}\right)-\tau\left(N_{c}-1-j\right)\right) 
	e^{\frac{i\pi }{2} \phi_2}
\end{align}

with:
\begin{align}
	\phi_2 =&
	-N_c(
	\hat{\lambda}^2 - (\hat{\lambda} + \hat{\nu}_0)	((N_c-1)\tau + 2\omega)
	+ \omega^2 + \frac{1}{2}(N_c - 1)\tau^2
	\nonumber\\&
	+N_{c}(
	4 \sigma^{\prime 2}-2 \mu_{1} \mu_{2}-2 \nu_{1} \nu_{2}-\left(N_{c}-1\right) \tau \sum_{\alpha=1,2}\left(\mu_{\alpha}+\nu_{\alpha}\right)
	\nonumber\\&
	\qquad 
	-\frac{2}{3}\left(N_{c}-1\right)\left(N_{c}-2\right) \tau^{2}
	)
	\nonumber\\
	=& \lambda ^2\frac{11 N_c}{144}+\frac{1}{72} \lambda  N_c \left(-9 \tau  (N_c-1)+36 \mu _1+36 \mu _2+14 \nu _1-86 \nu _2 -18 \omega \right) + \mathcal{O}(\lambda^0)
	\label{eq:exceptional_U0_U12_CT_lambda}
\end{align}

We notice that both in this duality and in the one between the $U(N_c)_0$ model and the $USp(2N_c)_2$ model there are contact terms for $\lambda^2$ with fractional coefficients \eqref{eq:exceptional_U0_U12_CT_lambda}, \eqref{eq:exceptional_U0_USp_CT_lambda}. These correspond to CS levels for the global topological symmetry $U(1)_J$ of the $U(N_c)_0$ theory. In order to perform our procedure we need to gauge a fraction of the topological symmetry in order to guarantee that after the gauging all the CS and mixed CS terms involving the (gauged) topological symmetry have integer level. Physically this implies that only for some values of $k_2$ the proposed duality makes sense. We find that the duality between the $U(N_c)_{0,k_2}$ theory and the $U(N_c)_{\frac{1}{2}}\times U(1) \times U(1)$ theory holds if $k_2$ is a multiple of $144$, while the duality between the $U(N_c)_{0,k_2}$ theory and the $USp(2N_c)_2\times U(1) \times U(1)$ theory holds if $k_2$ is a multiple of $4$. When both of these conditions are satisfied we have a duality between all the three phases described above. The integral identities among the three sphere partition functions of these three models are
\begin{align}
	Z_{U\left(N_{c}\right)_{0,144\hat{k_2}}}(\mu , \nu ; \tau ; \Lambda)=&
	\prod_{j=0}^{N_c-1}	\prod_{r=0}^{2}
	\Gamma_h (2\omega - (j \tau + \hat{\mu}_r + \hat{\nu}_1))
	\nonumber\\&
	\int d\xi dx
	e^{-\pi i  k_2 x^2+2 \pi i  \Lambda  x-2 \pi i  \xi  x}
	\quad  Z_{U(N_c)_{1/2}} (\hat{\mu}, \hat{\nu}; \tau;\hat{\lambda} )
	\nonumber\\&
	\prod_{j=0}^{N_{c}-1} \Gamma_{h}\Big(2 \omega \pm \frac{1}{2} \lambda-\frac{1}{2} \sum_{\alpha=1,2}(\mu_{\alpha}+\nu_{\alpha})-\tau(N_{c}-1-j)\Big) 
	e^{\frac{i\pi}{2}\phi_2 }
	\nonumber\\
	=&
	\int d\xi dx
	e^{-\pi i  k_2 x^2+2 \pi i  \Lambda  x-2 \pi i  \xi  x}
	\quad Z_{USp\left(2 N_{c}\right)_{2}}\left(\mu_{\sigma^{\prime}} ; \tau\right)
	\nonumber\\&
	\prod_{j=0}^{N_{c}-1} \Gamma_{h} \Big(2 \omega \pm \frac{1}{2} \lambda-\frac{1}{2} \sum_{\alpha=1,2}(\mu_{\alpha}+\nu_{\alpha})-\tau(N_{c}-1-j) \Big) 
	e^{\frac{i\pi}{2} \phi_1}
\end{align}
With $\xi= \lambda/24$ and $\phi_i$ given by eqs. \eqref{eq:exceptional_U0_U12_CT_lambda}, \eqref{eq:exceptional_Z_U0_USp_VandeBult}.

	Notice that in the phase with gauge group $U(N_c)_{\frac{1}{2}}$ the FI term $\hat{\lambda }$ corresponds to several mixed terms between the $U(N_c)_{\frac{1}{2}}$ non-abelian gauge group and the global symmetries. $\hat{\lambda}$ is fixed by the monopole superpotential to be a combination of the electric real masses and FI, in particular it contains the term:
	\begin{equation}
		5 (2 \pi i) \frac{\lambda}{24} \sum_{i=1}^{N_c} \hat{\sigma}_i 
	\end{equation}
	When $\xi= \frac{\lambda}{24}$ is gauged this is interpreted as a mixed CS term at level $-5$ between the $U(N_c)_{\frac{1}{2}}$ non-abelian gauge group and the abelian gauge sector $U(1)_J$.

%
%
%
%
%
%
%
%
%
%
%
%




%
%
%
%
%
%
\section{Further developments}
\label{sec:conc}
%
%
%
%
%
%

In this paper we have derived new three dimensional $\mathcal{N}=2$ dualities by a modification of the gauging/ungauging procedure of \cite{Aharony:2013dha}, consisting of an intermediate step, where we added a global CS  term for the new topological symmetry that arises when we gauge the original one.
This prescription allowed us to transform a $U(N_c)_{k_1} $ gauge group into  $U(N_c)_{k_1, k_1+N_c k_2}$. 
By modifying the dual phases  accordingly we have obtained new  
 dualities.
We have checked the  new dualities by matching the three sphere partition functions: by a careful treatment of  the original integral identities  we obtained the new integral identities for the new dualities.
We have studied in detail the case of SQCD and adjoint SQCD with non-chiral and chiral matter content. In the case of adjoint SQCD we have also discussed the case of a triality arising from the 
exceptional dualities of  \cite{Amariti:2018wht,Benvenuti:2018bav}.

Many generalizations of our work are possible.
Here we have checked our proposal by matching the three sphere partition function, but other checks should be necessary. For example one should match the computation of the
moduli space along the lines of \cite{Nii:2020ikd}. Furthermore one could also study the matching between the superconformal indices of the new dual pairs obtained here.
Then one can apply the prescription described here to other known dualities if they 
involve unitary gauge groups with an unbroken topological symmetry. For example the 3d version of  the \emph{Lagrangian}
Argyres-Douglas theory \cite{Benvenuti:2017lle,Benvenuti:2017kud,Aghaei:2017xqe,Nieri:2018pev} could be studied with our procedure.
Further interesting examples are the quiver gauge theories studied in \cite{Benvenuti:2020gvy,Benvenuti:2020wpc}.
Other  quivers that can be studied arise from the holographic correspondence,
as the ABJM theory \cite{Aharony:2008ug} and its non-chiral orbifolds. In this case it should be interesting to 
study the large $N$ scaling of the degrees of freedom along the lines of \cite{Herzog:2010hf,Jafferis:2011zi}.
One can also study examples with reduced supersymmetry $\mathcal{N}=1$ \cite{Benini:2018umh,Choi:2018ohn} or  
$\mathcal{N}=0$. In the last case some  $U(N_c)_{k_1, k_1+N_c k_2}$ gauge theories
 have been already investigated in \cite{Radicevic:2016wqn}, generalizing the duality of \cite{Aharony:2015mjs}.
 It should be interesting to apply the same generalization to the dualities of \cite{Benini:2017aed}.
Another important open question is related to the interpretation of the dualities
discussed here in terms of intersecting branes. Indeed Giveon-Kutasov and Aharony duality (and their generalizations to chiral  and/or  adjoint matter) have been engineered in terms
of branes \cite{Giveon:2008zn,Amariti:2015yea,Amariti:2015mva,Niarchos:2008jb,Amariti:2020xqm}
and  it is then natural to wonder if it should be possible to generalize such geometric constructions  to the cases with a different CS level for the abelian factors. 
In this case the primary difficulty stays in understanding the gauging/ungauging procedure on 
the brane picture. We hope to come back to this issue in the future.

%
%
%
%
%
%
\section*{Acknowledgments}
%
%
We are grateful to Marco Fazzi for discussion and for early collaboration on this project.
This work has been supported in part by the Italian Ministero dell'Istruzione, 
Universit\`a e Ricerca (MIUR), in part by Istituto Nazionale di Fisica Nucleare (INFN)
through the “Gauge Theories, Strings, Supergravity” (GSS) research project and in
part by MIUR-PRIN contract 2017CC72MK-003.  

\bibliographystyle{JHEP}
\bibliography{ref}

\end{document}